\documentclass[usenatbib,numberedappendix]{emulateapj}
\usepackage{mathrsfs}
\usepackage[colorlinks=true,linkcolor=black,citecolor=black,urlcolor=blue]{hyperref}
\usepackage{paralist}

\newcommand{\mpch}{\,h^{-1}{\rm {Mpc}}}

\newcommand{\lcdm}{\Lambda\mathrm{CDM}}

\newcommand{\bfv}{{\mathbf{v}}}

\newcommand{\bfr}{{\mathbf{r}}}

\newcommand{\bfs}{{\mathbf{s}}}

\newcommand{\rp}{{s_\perp}}
\newcommand{\rlos}{{s_\parallel}}
\newcommand{\kp}{{k_\perp}}
\newcommand{\klos}{{k_\parallel}}
\newcommand{\fig}{{Fig.\,}}
\newcommand{\seco}{{Sec.\,}}
\newcommand{\eqn}{{Eq.\,}}
\newcommand{\am}{{\mathrm{AM}}}
\newcommand{\amm}{{\mathrm{AM}^\mathrm{md}}}
\newcommand{\mpc}{{\mathrm{Mpc}}}
\newcommand{\gpc}{{\mathrm{Gpc}}}
\newcommand{\zeff}{{z_\mathrm{eff}}}

\begin{document}

\title{Measurement of Redshift Space Power Spectrum for BOSS
       galaxies and the Growth Rate at redshift 0.57}

\author{Zhigang Li\altaffilmark{1,2,*},
        Y.P. Jing\altaffilmark{1,2,*},
        Pengjie Zhang\altaffilmark{1,2,*},
        Dalong Cheng\altaffilmark{1} 
       }

\altaffiltext{1}{Center for Astronomy and Astrophysics, Department of
  Physics and Astronomy, Shanghai Jiao Tong University, 800 Dongchuan 
  Road, Shanghai, 200240}
\altaffiltext{2}{IFSA Collaborative Innovation Center, Shanghai Jiao Tong University, Shanghai 200240, China}
\altaffiltext{*}{zhigli@sjtu.edu.cn;ypjing@sjtu.edu.cn;zhangpj@sjtu.edu.cn}

\begin{abstract}
  We present a measurement of two-dimensional (2D) redshift-space
  power spectrum for the Baryon Oscillation Spectroscopic Survey
  (BOSS) Data Release 11 CMASS galaxies in the North Galactic Cap
  (NGC) based on the method developed by \cite{2001MNRAS.325.1389J}. 
  In this method, we first measure the 2D redshift-space correlation 
  function for the CMASS galaxies, and obtain the 2D power spectrum 
  based on Fourier Transform of the correlation function. The method 
  is tested with an N-body mock galaxy catalog, which demonstrates 
  that the method can yield an accurate and unbiased measurement of 
  the redshift-space power spectrum given the input 2D correlation
  function is correct. Compared with previous measurements in
  literature that are usually based on direct Fourier Transform in
  redshift space, our method has the advantages that the window
  function and shot-noise are fully corrected, while those measured in
  previous studies for the CMASS galaxies are usually the one
  convolved with the window function. In fact, our 2D power spectrum,
  by its construction, can accurately reproduce the 2D correlation
  function, and in the meanwhile can reproduce, for example, the 2D
  power spectrum of \cite{2014MNRAS.443.1065B} accurately if ours is convolved
  with the window function they provided. Thus, our measurement can
  facilitate a direct comparison with the theoretical predictions. With
  this accurate measurement of the 2D power spectrum, we then develop
  a method to measure the structure growth rate, by separating the
  anisotropic redshift-space power spectrum from the isotropic
  real-space power spectrum.  We have also carefully corrected for the
  nonlinearities in the mapping from real space to redshift space,
  according to the theoretical model of \cite{2013PhRvD..87f3526Z}. Finally, we
  obtain $f(\zeff)\sigma_8(\zeff)=0.438\pm0.037$ at the effective
  redshift $\zeff=0.57$, where $f(\zeff)$ is the linear growth rate
  and $\sigma_8(\zeff)$ is the rms density fluctuation in the
  sphere of comoving radius $8 \mpch$ at $\zeff$. The result
  is useful for constraining cosmological parameters. The measurements 
  of 2D power spectrum will be released soon.

\end{abstract}


\maketitle

%%%%%%%%%%%%%%%%%%%%%%%%%%%%%%%%%%%%%%%%%%%%%%%%%%%%%%%%%%%%%%%%%%%%%%%%%%%%%%%%
%%%%%%%%%%%%%%%%%%%%%%%%%%%%%%%%%%%%%%%%%%%%%%%%%%%%%%%%%%%%%%%%%%%%%%%%%%%%%%%%
%%%%%%%%%%%%%%%%%%%%%%%%%%%%%%%%%%%%%%%%%%%%%%%%%%%%%%%%%%%%%%%%%%%%%%%%%%%%%%%%
%%%%%%%%%%%%%%%%%%%%%%%%%%%%%%%%%%%%%%%%%%%%%%%%%%%%%%%%%%%%%%%%%%%%%%%%%%%%%%%%
%%%%%%%%   Section:  Introduction   %%%%%%%%%%%%%%%%%%%
%%%%%%%%%%%%%%%%%%%%%%%%%%%%%%%%%%%%%%%%%%%%%%%%%%%%%%%%%%%%%%%%%%%%%%%%%%%%%%%%
%%%%%%%%%%%%%%%%%%%%%%%%%%%%%%%%%%%%%%%%%%%%%%%%%%%%%%%%%%%%%%%%%%%%%%%%%%%%%%%%
%%%%%%%%%%%%%%%%%%%%%%%%%%%%%%%%%%%%%%%%%%%%%%%%%%%%%%%%%%%%%%%%%%%%%%%%%%%%%%%%
%%%%%%%%%%%%%%%%%%%%%%%%%%%%%%%%%%%%%%%%%%%%%%%%%%%%%%%%%%%%%%%%%%%%%%%%%%%%%%%%
\section{Introduction}      
\label{intro}
Redshift space distortion (RSD) is emerging as a major probe of
cosmology, and is playing an important role in ongoing and upcoming
dark energy surveys (e.g. \citet{Snowmasschapter4,stage5}). Peculiar
velocities of galaxies distort their distribution in redshift space
through the Doppler effect. They render the otherwise statistically
isotropic distribution of galaxies in real space into a statistically
anisotropic distribution in redshift space with a unique pattern.
Through such unique anisotropic pattern, in principle one is able to
infer statistical properties of peculiar velocities {\it at
  cosmological distances}. These statistics depend on both the law of
gravity, and the nature of gravitational sources (dark matter, dark
energy, etc.). It then provides us a precious tool to measure the
structure growth of the universe and to probe properties of dark
energy and gravity
(e.g. \citet{Amendola05,Yamamoto05,Jain08,Linder08,Wang08,Percival09,
  White09,Song09,Jennings11,Cai12}). Furthermore, the combination of
weak lensing and RSD allows for a test of General Relativity (GR),
insensitive to unknown galaxy bias and cosmic variances, through the
$E_G$ method
\citep{2007PhRvL..99n1302Z,2010Natur.464..256R,2015JCAP...12..051L,
  2015MNRAS.449.4326P,2015arXiv151104457P,2016MNRAS.456.2806B}.

A major challenge of RSD studies in the era of precision cosmology lies 
in its theoretical modelling, due to several nonlinear processes entangled 
in the redshift-space clustering of galaxies. However, precision measurement 
of RSD  also faces unresolved problems. RSD has been measured extensively 
using correlation function \citep{1993ApJ...417...19H,1994MNRAS.267..927F,
1996ApJ...468....1L,1998MNRAS.296..191R,2001Natur.410..169P,
2008Natur.451..541G,2008ApJ...676..889O,2009MNRAS.393.1183C,
2009MNRAS.396.1119C,2012MNRAS.423.3430B,2012MNRAS.426.2719R,
2013arXiv1312.4889C,2014MNRAS.444..476R,2014MNRAS.437.1109R,
2014MNRAS.439.3504S,2014MNRAS.440.2692S,2015MNRAS.449..848H,
2015MNRAS.453.1754A,2016PASJ...68...38O}. 
It has also been measured through the redshift-space power spectrum 
\citep{1994ApJ...431..569P,1995MNRAS.275..515C,1996ApJ...456L...1L,
2001MNRAS.325.1389J,2004ApJ...617..782J,2006PASJ...58...93Y,
2008PThPh.120..609Y,2010PhRvD..81j3517Y,2011MNRAS.415.2876B,
2013JCAP...08..019H,2014MNRAS.439.2515O,2014MNRAS.443.1065B,
2015PhRvD..92b3523K,2016MNRAS.tmp..927G,2016arXiv160600439G,
2016MNRAS.458.2725J}
and bispectrum \citep{2015MNRAS.451..539G,2015MNRAS.452.1914G,
2016arXiv160600439G}.  The redshift-space power spectrum is more
directly connected to the theory of large scale structure (LSS). 
However, its precision measurement faces two difficulties. 

One problem is that RSD effect is along different line-of-sights (LOS) 
for different galaxies, while Fourier transform tends to mix different 
LOSs.  In early works, the power spectrum analysis was usually based 
on the parallel-plane approximation, that is, all galaxies in the 
survey share one unique LOS direction. Then one can rely on the Fast 
Fourier Transform (FFT) technique to accelerate the power spectrum 
calculation. However, the sky area covered by galaxy surveys becomes 
larger and larger. The parallel-plane approximation becomes
less and less accurate and systematics introduced by the variation 
of LOSs in the survey becomes non-negligible.  One solution beyond the 
parallel-plane approximation was proposed by \cite{2006PASJ...58...93Y} 
(Y06 hereafter). It has been implemented in various recent galaxy surveys  
\citep{2006PASJ...58...93Y, 2008PThPh.120..609Y, 2010PhRvD..81j3517Y,
2013JCAP...08..019H, 2014MNRAS.439.2515O, 2014MNRAS.443.1065B,
2015PhRvD..92b3523K, 2016MNRAS.tmp..927G, 2016arXiv160600439G,
2016MNRAS.458.2725J}.  In the Y06 method, each galaxy pair shares a 
common LOS, which is further approximated as that of one galaxy in 
the pair. With this `moving-LOS approximation', the pair summation 
can be implemented by two Fourier transforms and then can be accelerated 
by FFT. Nevertheless, this may introduce notable systematics on the 
hexadecapole power  spectrum for wide galaxy surveys 
\citep{2012MNRAS.420.2102S,2015MNRAS.447.1789Y}.  Another proposed
solution is to decompose the 3D density field with the spherical 
harmonics and spherical Bessel functions, and is referred as 
SFB\footnote{In some literatures, it is called the spherical 
Fourier-Bessel expansion (SFB for short).} hereafter
\citep{1995MNRAS.275..483H,1995MNRAS.272..885F,1999MNRAS.305..527T,
2000MNRAS.317L..23H,2001MNRAS.327..689T,2002MNRAS.335..887T,
2004ApJ...606..702T,2004MNRAS.353.1201P,2006MNRAS.373...45E,
2012A&A...540A..92L,2012A&A...540A..60L,2012A&A...540A.115R,
2015A&A...578A..10L}. This decomposition keeps the LOS information
(and therefore RSD information) exactly.  However, the measured SFB 
power spectrum differs from the redshift-space power spectrum predicted 
by most commonly used RSD models. Furthermore, it mixes clustering at 
different redshifts. Both bring inconveniences when one compares the 
data with models.

Another important issue in the power spectrum measurements is the 
deconvolution of window function. Unlike that in the correlation 
function, it is nontrivial to correct the window function in Fourier 
space. On one hand, the window function couples different Fourier 
modes. On the other hand, the window function introduces non-uniform 
distribution of $\mu=\hat{k}\cdot\hat{n}_\mathrm{LOS}$. Here, 
$\hat{k}$ is  the unit wavenumber vector and $\hat{n}_\mathrm{LOS}$ 
is the unit LOS vector. Such non-uniform $\mu$ distribution may bias 
the measurement of power spectrum multipoles 
\citep{2012MNRAS.420.2102S,2015MNRAS.447.1789Y}.

In this paper, we propose to use the two-dimensional (2D) galaxy power 
spectrum in redshift space to measure the RSD effect, instead of using 
the multipole power spectrum or SFB coefficients. We revisit the method 
of measuring the 2D galaxy power spectrum through Fourier transform of 
the 2D galaxy correlation function in redshift space developed by 
\cite{2001MNRAS.325.1389J}. This method improves the parallel-pane 
approximation and the `moving-LOS' approximation. In measurement of 2D 
correlation function, LOS is defined on each galaxy pair, usually to 
be the position vector of pair center with respect to the observer. 
The LOS defined in this way captures all information of RSD under the 
assumption of distant observer and neglecting wide-angle effect. 
Furthermore, the window function can be corrected in configuration 
space robustly and efficiently since the deconvolution in Fourier 
space becomes division in configuration space. The non-uniform 
$\mu$-distribution can also be solved by uniformly weighting the 
correlation function in $(s,\mu)$ space.

Next, we propose a method to measure the structure growth rate through
the 2D power spectrum measurement. We separate the anisotropies on the
galaxy power spectrum in redshift space from the isotropic galaxy
power spectrum in real space by introducing a new statistics -
anisotropic measure. When modeling the anisotropic measure, we have
corrected for the nonlinearities with the theoretical model of Zhang
et al (2013). In this way, the RSD parameter and galaxy bias can be
measured independently. We apply the method to the BOSS-DR11 CMASS
galaxy sample and obtain a robust measurement of the structure growth
rate.

The paper is organized as follows. In \S 2, we introduce the method of
measuring the 2D power spectrum in redshift space for a large galaxy
survey and test it with a mock galaxy catalog based on an $N$-body
simulation.  In \S 3, we introduce the data set used in this paper:
BOSS-DR11 CMASS galaxy sample and the MD-Patchy mock galaxy
catalogs. In \S 4, we show the measured 2D power spectrum of
BOSS-DR11 CMASS galaxies. We measure the structure growth rate from
the measured 2D power spectrum and compare it with previous studies in
\S 5. We end the paper with a brief summary in \S 6.

%%%%%%%%%%%%%%%%%%%%%%%%%%%%%%%%%%%%%%%%%%%%%%%%%%%%%%%%%%%%%%%%%%%%%%%%%%%%%%%%
%%%%%%%%%%%%%%%%%%%%%%%%%%%%%%%%%%%%%%%%%%%%%%%%%%%%%%%%%%%%%%%%%%%%%%%%%%%%%%%%
%%%%%%%%%%%%%%%%%%%%%%%%%%%%%%%%%%%%%%%%%%%%%%%%%%%%%%%%%%%%%%%%%%%%%%%%%%%%%%%%
%%%%%%%%%%%%%%%%%%%%%%%%%%%%%%%%%%%%%%%%%%%%%%%%%%%%%%%%%%%%%%%%%%%%%%%%%%%%%%%%
%%%%%%%%   Section:  Method   %%%%%%%%%%%%%%%%%%%
%%%%%%%%%%%%%%%%%%%%%%%%%%%%%%%%%%%%%%%%%%%%%%%%%%%%%%%%%%%%%%%%%%%%%%%%%%%%%%%%
%%%%%%%%%%%%%%%%%%%%%%%%%%%%%%%%%%%%%%%%%%%%%%%%%%%%%%%%%%%%%%%%%%%%%%%%%%%%%%%%
%%%%%%%%%%%%%%%%%%%%%%%%%%%%%%%%%%%%%%%%%%%%%%%%%%%%%%%%%%%%%%%%%%%%%%%%%%%%%%%%
%%%%%%%%%%%%%%%%%%%%%%%%%%%%%%%%%%%%%%%%%%%%%%%%%%%%%%%%%%%%%%%%%%%%%%%%%%%%%%%%
\section{Method of Measuring Redshift Space Distortion Power Spectrum}  
\label{Sec:method}
In this section, we describe our method to measure the 2D power
spectrum from a large redshift survey. We will also verify this method
against the mock galaxy distribution in a high-resolution simulation.

Peculiar velocity $\bfv$ of a galaxy adds a Doppler redshift on top
of its cosmological redshift. Therefore,  the real-space position 
at $\bfr$ changes to the corresponding position at
$\bfs_1$ in redshift-space
\begin{equation}
\label{EQ:Method:rs0}
 \bfs_1=\bfr+\left[\frac{\bfv\cdot \hat{r}}{H(z)}\right]\hat{r}\ .
\end{equation}
Each galaxy has its own LOS ($\hat{r}$) and therefore only the
velocity component (${\bf v}\cdot\hat{r}$) along the LOS contributes to RSD. Large
surveys can have very different LOSs, so the variation of LOSs must 
be taken into account. A direct Fourier transform mixes all LOSs in 
the survey volume and therefore can erase most, if not all, RSD signal. 
In correlation function, this problem is much suppressed. For a pair of 
galaxies at redshift-space positions $\bfs_{1}$ and $\bfs_{2}$,, we can decompose 
the separation vector $\bfs\equiv \bfs_2-\bfs_1$  into 
$\bfs=(\bfs_{\perp}, s_\parallel)$, where $s_\parallel$ is the
separation along the LOS pointing to the center of the pair 
${\bf s}_h\equiv ({\bf s}_1+{\bf s}_2)/2$ (that is 
$s_\parallel=\bfs\cdot \hat{\bfs}_h$) and $\bfs_\perp$ is the 2D 
transverse component of $\bfs$. We can then measure the correlation 
function $\xi^s({\bf s}_\perp,s_{\parallel})$. The superscript `$s$'
denotes redshift-space property. From symmetry argument, the 
correlation function depends on the amplitude of $\bfs_\perp$, but 
not its direction. Then the expectation value of the correlation 
function depends only on $s_\perp$ and $s_{\parallel}$. Therefore we 
often call $\xi^s(s_\perp,s_{\parallel})$ the 2D correlation function, 
or the anisotropic correlation function. Instead of approximating all 
LOSs as a single LOS in the direct Fourier transform, the correlation 
function measurement only requires that the two LOSs ($\bfs_{1,2}$) of a 
given pair can be approximated as the LOS  to the pair center ($\bfs_h$). 
Therefore it is a much more accurate approximation. The accuracy is of
the order $\theta^2/2=1.5\% (\theta/10^{\circ})^2$ where $\theta$ is 
the angular separation of the pair. For BOSS CMASS galaxies we analyze 
($z_\mathrm{med}=0.57$) and for the scale $s_\perp \la 300\mpch$ we 
are interested, the accuracy is better than $2\%$. So we can neglect 
the error caused by this approximation\footnote{The wide-angle effect 
has been shown to be small in SDSS-like galaxy surveys and is expected 
be even smaller in the BOSS-like galaxy surveys 
\citep{2012MNRAS.420.2102S,2015MNRAS.447.1789Y}}. Therefore the 
correlation function $\xi^s(s_{\perp},s_\parallel)$ faithfully 
captures the RSD effect. 

Nevertheless, from the viewpoint of theoretical modeling and 
cosmological parameter fitting, the 2D power spectrum is more 
convenient. First, it is more directly connected to the theory of 
LSS. Second,  it is more straightforward to cut in $(k,\mu)$ space 
to minimize uncertainties of cosmological parameter fitting,  arising 
from various nonlinearities. Nonlinearities in real-space  clustering 
can be mitigated by cut in $k$, while nonlinearities in real 
space-redshift space mapping can be mitigated by cut in $\mu$. The 
major goal of this paper is to test the method of measuring the 2D 
power spectrum and apply it on the BOSS-DR11 CMASS galaxies.

%%%%%%%%%%%%%%%%%%%%%%%%%%%%%%%%%%%%%%%%%%%55%%%%%%%%%%%%%%%%%
%%%%%   Subsection: measurement of 2D power spectrum 
%%%%%%%%%%%%%%%%%%%%%%%%%%%%%%%%%%%%%%%%%%%55%%%%%%%%%%%%%%%%%
%%%%%%%%%%%%%%%%%%%%%%%%%%%%%%%%%%%%%%%%%%%%%%%%%%%%%%%%%%%%%%
\subsection{Measurement of 2D Power Spectrum}
\label{Sec:Method:pipeline}

Following \cite{2001MNRAS.325.1389J}, we apply a correlation function 
based method to measure the anisotropic redshift-space power spectrum 
of BOSS galaxies.  It is a two-step procedure. We first measure the 
two-dimensional correlation function $\xi^s(s_\perp,s_\parallel)$. 
Then we translate the 2D correlation function to the 2D power spectrum 
by Fourier transform. 

%%%%%%%%%%%%%%%%%%%%%%%%%%%%%%%%%%%%%%%%%%%%%%%%%%%
%%%%%   Subsubsection: 2D correlation function
%%%%%%%%%%%%%%%%%%%%%%%%%%%%%%%%%%%%%%%%%%%%%%%%%%%
\subsubsection{Measuring the 2D correlation function}
\label{Sec:Method:2dcf}
The redshift-space galaxy correlation function is measured using the
Landy-Szalay estimator,
\begin{equation}
\xi^s_g(s,\mu_s) = \frac{\mathrm{DD-2DR+RR}}{\mathrm{RR}}\ .
\label{EQ:Method:ls}
\end{equation}
Notice that, instead of binning in $s_\perp$-$s_\parallel$ space, we
bin in $s$-$\mu_s$ space. Here, 
$s\equiv \sqrt{s_{\perp}^2+s_{\parallel}^2}$ and $\mu_s=s_\parallel/s$. 
DD represents the normalized number of galaxy-galaxy pairs 
whose separation lies in the corresponding $(s,\mu_s)$ bins, DR 
the number of galaxy-random pairs and RR the number of random-random 
pairs. The normalizations are taken as the total number of pairs for 
each component, i.e. $N_g(N_g-1)/2$ for DD, $N_gN_R$ for DR and 
$N_R(N_R-1)/2$ for RR with total number of galaxies $N_g$ and total 
number of random points $N_R$. The number of random points are taken 
to be about 100 times of the number of galaxies which is large enough 
for the correction of window function. 

We use adaptive bin size in $s$-direction to reach high resolution on 
small scales and maintain reasonable signal-to-noise ratio on large 
scales simultaneously. We have a total of 36 bins in the range of 
$0<s<300\mpch$: two bins are in the range of $0-2\mpch$; 3 bins in 
$2-8\mpch$; 4 bins in $8-24\mpch$; 17 bins in $24-160\mpch$; 10 bins 
in $160-300\mpch$. The bin sizes within each range are equal, which 
are $1\mpch$, $2\mpch$, $4\mpch$, $8\mpch$ and $14\mpch$, respectively.

In the $\mu_s$-direction, we use 20 bins with equal size of $\Delta\mu_s=0.05$.
We have checked that differences of binning in $(s,\mu_s)$ space result 
in negligible changes in the measured power spectrum. In the end, we 
interpolate the resulting 2D correlation function $\xi^s(s,\mu_s)$ on 
fine grids of $(\rp,\rlos)$ space for calculation of the 2D power 
spectrum in the next step, where $\rlos=s\cdot\mu_s$ and 
$\rp=\sqrt{s^2-\rlos^2}$.

%%%%%%%%%%%%%%%%%%%%%%%%%%%%%%%%%%%%%%%%%%%%%%%%%%%
%%%%%   Subsubsection: 2D power spectrum
%%%%%%%%%%%%%%%%%%%%%%%%%%%%%%%%%%%%%%%%%%%%%%%%%%%
\subsubsection{Measuring the 2D power spectrum}
\label{Sec:Method:2dps}
From the 2D correlation function, we calculate the 2D power spectrum 
using the Fourier transform,
\begin{eqnarray}
\label{EQ:Method:Pkmu}
 P_g^s(k,\mu) &=&\int \xi_g^s(\rp,\rlos) e^{i{\bf
     k}\cdot{\bf s}}d^3{\bf s}\nonumber \\
&=& \int \xi_g^s(\rp,\rlos) 
    e^{i(\klos\rlos+\kp\rp\cos(\phi))} 
    \rp d\rp d\phi d\rlos \nonumber \\
   &=& \int \xi_g^s(\rp,\rlos) K(\kp,\klos;\rp,\rlos)\rp 
       d\rp d\rlos\ ,
\end{eqnarray}
where $\klos=k\cdot\mu$ and $\kp=\sqrt{k^2-\klos^2}$. Notice that 
$\mu$ here is not related to  $\mu_s$ in the correlation function. 
The kernel $K$ is defined as 
$K(\kp,\klos;\rp,\rlos) = \cos(\klos\rlos) J_0(\kp\rp)$ with  
$J_0(x)=\int e^{ix\cos(\phi)} d\phi$ the zero-th order Bessel function. 
In practice, we need to cut the integral at some maximum value 
$s_\mathrm{max}$ to avoid contaminations from poor data at large $s$. 
But too small a value of $s_\mathrm{max}$ will introduce significant bias 
on the power spectrum, i.e. suppress power on large scales. For 
BOSS-DR11 CMASS galaxy sample that we will analyze later, we have 
tested that $s_\mathrm{max}=300\,h^{-1}\mathrm{Mpc}$ is an appropriate 
choice without introducing notable bias. The computational cost to 
calculate 2D correlation function up to this $s_\mathrm{max}$ is 
about 48 hours on a workstation for BOSS-DR11 CMASS NGC galaxies 
with 100 times more random points. For mock samples, we use 10 times 
more random points than mock galaxies. So we only need 24 CPU hours 
to measure 2D correlation function for one mock sample. 

The multipoles of 2D power spectrum can be calculated by
\begin{equation}
 P^s_{g,l}(k) = (2l+1) \int_0^1 P_g^s(k,\mu) 
    \mathcal{L}_l(\mu) d\mu\ .
\label{EQ:Method:pkl}
\end{equation}
The first few Legendre polynomials we used are, $\mathcal{L}_0(x)=1$, 
$\mathcal{L}_2(x)=(3x^2-1)/2$ and $\mathcal{L}_4(x)=(35x^4-30x^2+3)/8$.

%%%%%%%%%%%%%%%%%%%%%%%%%%%%%%%%%%%%%%%%%%%%%%%%%%%
%%%%%   Subsubsection: Test on simulation
%%%%%%%%%%%%%%%%%%%%%%%%%%%%%%%%%%%%%%%%%%%%%%%%%%%
\subsection{Test with an N-body simulation}
\label{Sec:Method:sim}

\begin{figure*}
\centering
{\includegraphics[width=6.5in]{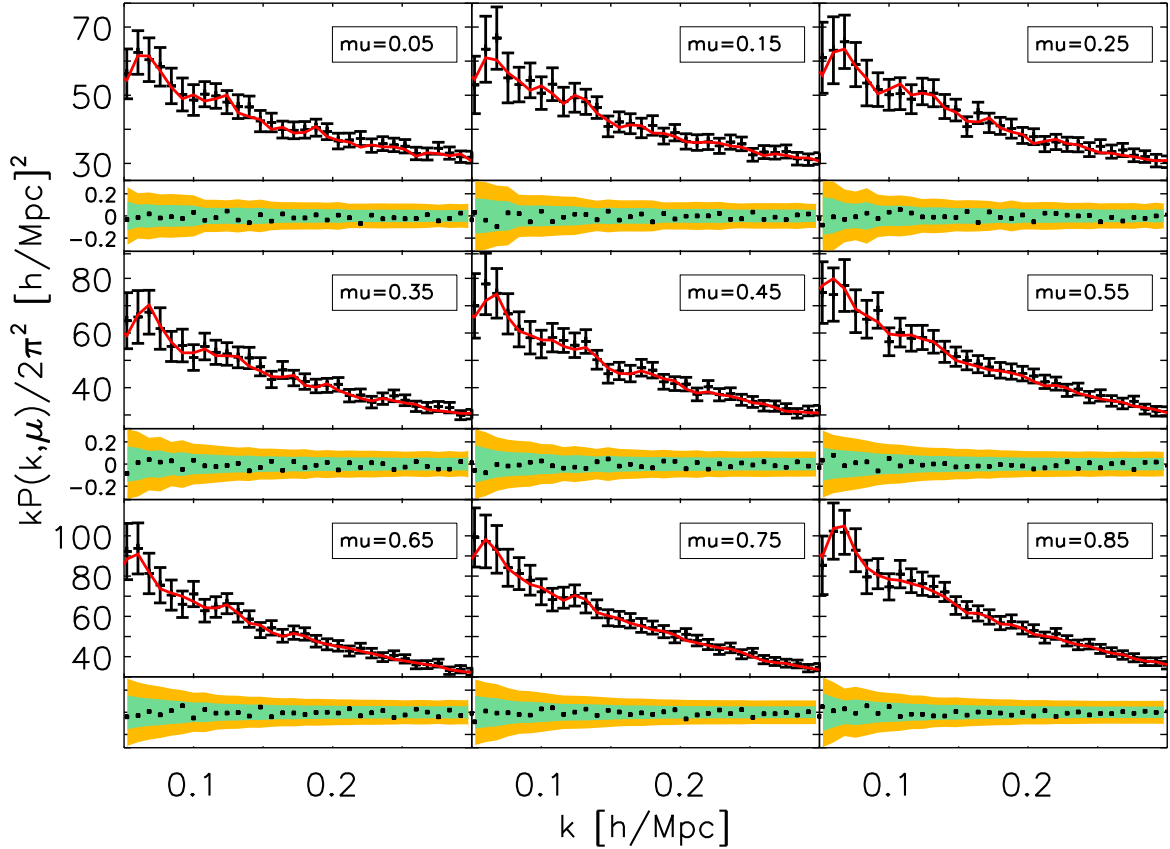}}
\caption{2D power spectrum of mock galaxies in the 
CosmicGrowth simulation. The nine panels show $kP(k,\mu)/2\pi^2$ 
distribution as a function of $k$, for specific $\mu$ bins. The power 
spectrum we obtained with our method are shown as red solid lines and 
those using FFT method are shown as black plus signs in the top 
sub-box of each panel. The error bars for the data points of FFT 
method are estimated according to \eqn\ref{EQ:method:error}. 
In the bottom sub-box of each panel shows the difference between the 
two methods, $(P(k,\mu)-P_\mathrm{FFT}(k,\mu))/P_\mathrm{FFT}(k,\mu)$. 
The Celadon green (Chrome yellow) band shows $1\,\sigma$ ($2\,\sigma$) 
confidence level. The two methods show good consistency. The differences 
between them are well within the $1\,\sigma$ level.}
\label{fig01}
\end{figure*}

We test our redshift-space power spectrum measurements using mock galaxies 
constructed from an $N$-body simulation, where we can build redshift-space 
mock galaxy distribution without wide-angle effect. This allows us to make 
an exact comparison between our method and the FFT method.

The simulation we used is one of high-resolution CosmicGrowth simulations 
(Jing, in preparation), which were generated by the P$^3$M code of
\cite{2007ApJ...657..664J} with $3072^3$ particles contained in a box
of 1.2 $h^{-1}\gpc$ on a side. The initial condition is made at
redshift $z_i=144$ following the Zeldovich approximation with the
transfer function from \cite{1996ApJ...469..437S}. The cosmological
parameters are set as $\Omega_m=0.268$, $\Omega_b=0.044$,
$\Omega_\Lambda=0.732$, $n_s=0.96$ and $\sigma_8 = 0.83$. 
The simulation run 5000 even steps in the expansion factor to 
redshift $z=0$. We further made the host halo catalog through 
the Friend-Of-Friend (FOF) method with a link parameter
$b_\mathrm{link}=0.2$. The sub-halos are identified 
using the Hierarchical-Bound-Tracing (HBT) algorithm
\citep{2012MNRAS.427.2437H}.

We relate galaxies to sub-halos and infer the stellar mass for 
galaxies using the subhalo-galaxy matching method proposed by 
\cite{2006MNRAS.371..537W}, 
\begin{eqnarray}
M_\mathrm{star}=\frac{2c}{(M_\mathrm{infall}/M_0)^{-a}+
  (M_\mathrm{infall}/M_0)^{-b}}
\end{eqnarray}
where the infall mass $M_\mathrm{infall}$ is defined as the mass of
the sub-halo at the time when it was last the central dominant object.
Then we introduce a Gaussian scatter on $\log(M_\mathrm{star})$ at a
given value of $M_\mathrm{infall}$ with dispersion $\sigma_M$. We
adopt the parameters given in \cite{2010MNRAS.402.1796W} which can fit
the stellar mass function and correlation function of SDSS galaxies at
$z=0$ and VVDS galaxies at $z=0.83$ simultaneously: $a=0.29$,
$b=2.42$, $c=10^{10.15}\,h^{-1}M_\odot$,
$M_0=4.34\times10^{11}\,h^{-1}M_\odot$ and $\sigma_M=0.24$.  We use a
stellar mass threshold of $10^{10.98}\,M_\odot$ to obtain number
density of $\bar{n}_\mathrm{gal}=4\times10^{-4}\,(\mpc/h)^{-3}$ (which
is the observed peak density at $z=0.5$) for mock galaxies. The mock
galaxy catalog has a bias of $b_\mathrm{gal}=1.68$.

In order to convert the real-space position (box coordinates) to 
redshift-space coordinate, we assume a distant observer in the 
$\hat{\mathbf{z}}$ direction, obtaining $s=r+v_z/aH$.  Where $v_z$ 
is the physical peculiar velocity along $z$-direction and the value 
of $aH$ is taken at redshift $z=0.57$. To enhance the S/N ratio, this 
procedure is also applied to the $\hat{\mathbf{x}}$ and $\hat{\mathbf{y}}$ 
directions. In the end, power spectrum is averaged over those obtained 
for each specific direction (observer). 

Next, we measure the 2D power spectrum in redshift space for the 
resulting mock galaxy catalog, following the same procedure described in 
\seco\ref{Sec:Method:pipeline}. We also use the FFT method to measure 
the redshift-space 2D power spectrum for the same mock galaxy catalog. 
The size of the FFT box is $1200\mpch$ at one side. We use $1200^3$ 
grids on which the galaxy densities are calculated and Fourier 
transformed to get the power spectrum. For the FFT method, we estimate 
the errors of the 2D power spectrum by assuming a Gaussian distribution, 
\begin{eqnarray}
\label{EQ:method:error}
\sigma_P=\sqrt{\frac{2}{N_\mathrm{mode}(k,\mu)}} 
 (P^s(k,\mu)+\frac{1}{\bar{n}_\mathrm{gal}})
\end{eqnarray}
where $N_\mathrm{mode}(k,\mu)$ is the number of modes residing in the 
$(k,\mu)$-bin, $\bar{n}_\mathrm{gal}$ is the mean number density of 
mock galaxies and $P^s(k,\mu)$ is the power spectrum of mock galaxies. 
The first term in the right-hand side is the cosmic variance and the 
second term is shot-noise.

The measured 2D power spectrums of mock galaxies are shown in 
\fig\ref{fig01}. The 9 panels show $kP^s(k,\mu)/2\pi^2$ 
for 9 $\mu$-bins. We also show the difference between the two methods,
$(P^s(k,\mu)-P^s_\mathrm{FFT}(k,\mu))/P^s_\mathrm{FFT}(k,\mu)$, in the
bottom sub-box in each panel. Where $P^s(k,\mu)$ stands for the 2D
power spectrum measured using our method and $P^s_\mathrm{FFT}(k,\mu)$
for the direct FFT method. The two methods show good consistency over
almost all scales. The differences between them are well within
$1\,\sigma$ level. Although they are omitted, the 2D power spectrums
at $\mu=0.95$ bin are consistent at the same level as
others. Therefore we verify that obtaining the 2D power spectrum from
the 2D correlation function is unbiased.  Together with the
sophisticated and accurate method of measuring the 2D correlation
function, we conclude that the 2D power spectrum measured in this way
can faithfully capture the RSD effect.

Furthermore, the 2D power spectrum  based on the Fourier transform of 
the2D correlation function has a few advantages: 
\begin{itemize}
\item The 2D power spectrum is free of normalization and shot-noise 
subtraction, since the correlation function is free of such issues. 
\item The survey window function is dealt with in configuration 
space when measuring the 2D correlation function, thus avoiding the 
deconvolution problem in the traditional multipole power spectrum 
measurements. It has been shown that the decoupling of window 
function in configuration space is stable and efficient.
\item The nonuniform distribution of cosine angle $\mu$ can be solved 
when measuring 2D power spectrum multipoles (\eqn\ref{EQ:Method:pkl}). 
The main difference between our method and the traditional multipole 
power spectrum measurements is the way of weighting the data. In the 
traditional multipole power spectrum measurements, each galaxy or 
galaxy pair has equal weight. However, the $\mu$-distribution of 
galaxy pairs is commonly nonuniform in realistic galaxy surveys and 
so introduces systematics on the measured multipole power spectrum. 
We apply equal weight to the 2D correlation function where the 
$\mu$-distribution can be sufficiently uniform. It is worth noting
that the method of Y06 without the `moving-LOS approximation' is 
formally equivalent to the method used in this paper to measure the 
multipole power spectrum. 
\item The wide-angle effect can be reduced. The 2D correlation function 
at large $s$-scales usually has low signal-to-noise ratio and contains 
little cosmological information. When calculating the 2D power spectrum 
from the 2D correlation function, we would cut the integral at some 
maximum value of $s_\mathrm{max}$ to prevent the contamination of poor 
data. This effectively reduces the impact of wide-angle effect which is 
important only at large separations. 
\end{itemize}

%%%%%%%%%%%%%%%%%%%%%%%%%%%%%%%%%%%%%%%%%%%%%%%%%%%%%%%%%%%%%%%%%%%%%%%%%%%%%%%%
%%%%%%%%%%%%%%%%%%%%%%%%%%%%%%%%%%%%%%%%%%%%%%%%%%%%%%%%%%%%%%%%%%%%%%%%%%%%%%%%
%%%%%%%%%%%%%%%%%%%%%%%%%%%%%%%%%%%%%%%%%%%%%%%%%%%%%%%%%%%%%%%%%%%%%%%%%%%%%%%%
%%%%%%%%%%%%%%%%%%%%%%%%%%%%%%%%%%%%%%%%%%%%%%%%%%%%%%%%%%%%%%%%%%%%%%%%%%%%%%%%
%%%%%%%%  Section:  Data Set   %%%%%%%%%%%%%%%%%%%
%%%%%%%%%%%%%%%%%%%%%%%%%%%%%%%%%%%%%%%%%%%%%%%%%%%%%%%%%%%%%%%%%%%%%%%%%%%%%%%%
%%%%%%%%%%%%%%%%%%%%%%%%%%%%%%%%%%%%%%%%%%%%%%%%%%%%%%%%%%%%%%%%%%%%%%%%%%%%%%%%
%%%%%%%%%%%%%%%%%%%%%%%%%%%%%%%%%%%%%%%%%%%%%%%%%%%%%%%%%%%%%%%%%%%%%%%%%%%%%%%%
%%%%%%%%%%%%%%%%%%%%%%%%%%%%%%%%%%%%%%%%%%%%%%%%%%%%%%%%%%%%%%%%%%%%%%%%%%%%%%%%
\section{Data Set}
\label{Sec:Data}

%%%%%%%%%%%%%%%%%%%%%%%%%%%%%%%%%%%%%%%%%%%%%%%%%%%
%%%%%%%%%%%%%%%%%%%%%%%%%%%%%%%%%%%%%%%%%%%%%%%%%%%
%%%%%   Subsection: Galaxies
%%%%%%%%%%%%%%%%%%%%%%%%%%%%%%%%%%%%%%%%%%%%%%%%%%%
%%%%%%%%%%%%%%%%%%%%%%%%%%%%%%%%%%%%%%%%%%%%%%%%%%%
\subsection {BOSS-DR11 CMASS Galaxies}
\label{Sec:Data:cmass}

In this paper, we use the publicly released CMASS galaxy sample in 
the Data Release 11 of Sloan Digital Sky Survey III (SDSS III) Baryon 
Oscillations Spectroscopic Survey (BOSS). SDSS has scanned over one 
third of the sky and obtained images in five photometric bandpasses 
to a limiting magnitude of $r_\mathrm{lim}\approx22.5$ using the 2.5 
meter Sloan Telescope located at Apache Point Observatory in New 
Mexico. As a part of SDSS III, BOSS is designed to do spectroscopic 
observations of more than one million galaxies covering 10000 square 
degrees on the sky. Within BOSS, the CMASS sample is approximately 
stellar-mass limited above $z=0.45$, and target galaxies are selected 
from SDSS DR8 imaging data with selection function described in 
\cite{2014MNRAS.441...24A}. The median redshift of CMASS galaxies 
is at $z_\mathrm{med}\approx0.57$ and the stellar mass peaks at 
$M_\mathrm{stellar}\approx10^{11.3}\,M_\odot$. Most of the CMASS 
galaxies are central galaxies in dark matter halos of mass about 
$10^{13}\,h^{-1}M_\odot$ with a non-negligible fraction of satellites 
which reside in halos about 10 times more massive. The BOSS-DR11 CMASS 
sample contains 690,826 galaxies and covers 8498 square degrees. We 
limit our analysis to the North Galactic Cap (NGC) sample which contains 
520,805 galaxies and covers 6769 square degrees. 

We correct for the effects of redshift failure and fiber collision by 
up-weighting the galaxies whose nearest neighbor had a redshift failure 
($w_\mathrm{rf}$) or failed to get redshift because they are a close pair 
($w_\mathrm{cp}$). We also apply the systematic weights to account for 
the seeing effect ($w_\mathrm{seeing}$) and correlation between the 
number density of observed galaxies and the stellar density 
($w_\mathrm{star}$). All of these weights are documented in the publicly 
released data. To reach minimum variance for galaxy clustering measurement,  
we apply the FKP weight \citep{1994ApJ...426...23F} in a simple form, 
$w_\mathrm{FKP}(\mathbf{r})=1/(1+\bar{n}(\mathbf{r})P_0)$ with
$P_0=20000$, where $\bar{n}(\mathbf{r})$ is the expected galaxy number density. 
The total weight applied to each galaxy is then, 
\begin{eqnarray}
w_\mathrm{tot}=(w_\mathrm{rf}+w_\mathrm{cp}-1)w_\mathrm{seeing}
  w_\mathrm{star}w_\mathrm{FKP}
\end{eqnarray}
The survey completeness has been carefully calculated and publicly
released together with the catalog. The random points are generated
following the survey completeness in light of the `Mangle' software
\citep{2004MNRAS.349..115H,2008MNRAS.387.1391S}.  The redshifts and
FKP weights are `shuffled' (e.g., randomly selected) from the
observational sample to be assigned to the random points following a
uniform distribution \citep{2012MNRAS.424..564R}.  Please refer to
\cite{2014MNRAS.441...24A} for more details about the BOSS-DR11 CMASS
galaxy sample and \cite{2012MNRAS.424..564R} for the effect of the
various weights.

For this observational sample, we assume a $\lcdm$ cosmology to 
transfer the redshift to comoving distance with parameters, 
$\Omega^\mathrm{fid}_m=0.274$, $\Omega^\mathrm{fid}_\Lambda=0.726$.

%%%%%%%%%%%%%%%%%%%%%%%%%%%%%%%%%%%%%%%%%%%%%%%%%%%
%%%%%%%%%%%%%%%%%%%%%%%%%%%%%%%%%%%%%%%%%%%%%%%%%%%
%%%%%   Subsection: Mock catalog 
%%%%%%%%%%%%%%%%%%%%%%%%%%%%%%%%%%%%%%%%%%%%%%%%%%%
%%%%%%%%%%%%%%%%%%%%%%%%%%%%%%%%%%%%%%%%%%%%%%%%%%%
\subsection {Mock Catalogs to measure the covariance matrix}
\label{Sec:Data:mock}
To calculate covariance matrix for the 2D power spectrum and
anisotropic measure defined in the following section, we use the
MultiDark Patchy mock catalogs (hereafter MD-Patchy mocks) for BOSS
DR11 CMASS sample \citep{2016MNRAS.456.4156K}.  The MD-Patchy mocks
are constructed relying on the PATCHY approximate simulations of dark
matter density fields and using a biasing model to populate galaxies
in the dark matter density fields
\citep{2013MNRAS.435L..78K,2014MNRAS.439L..21K,2015MNRAS.450.1836K}.
A coherent peculiar velocity field is calculated using the augmented
Lagrangian Perturbation Theory (ALPT) consistently with the
displacement field. The finger-of-god (FOG) effect is modeled using a
Gaussian distribution function with parameters calibrated on the
monopole and quadrupole damping effect in BigMultiDark simulation and
BOSS observational data. The MD-Patchy mock catalogs are constructed
assuming a $\lcdm$ Planck cosmology: $\Omega^\mathrm{MD}_m=0.307115$,
$\Omega^\mathrm{MD}_b=0.048206$, $\sigma^\mathrm{MD}_8=0.8288$,
$n^\mathrm{MD}_s=0.9611$ and $h^\mathrm{MD}=0.6777$. The resulting
MD-Patchy mocks reproduce the number density, selection function,
survey geometry, multipole power spectrum, multipole correlation
function and three point statistics of the BOSS DR11 CMASS sample. The
MD-Patchy mock catalogs have been tested and applied on the analysis
of BOSS galaxy surveys
\citep{2013arXiv1312.4889C,2016arXiv160505352C,2016MNRAS.457.1770C}.

\begin{figure}
\resizebox{\hsize}{!}
{\includegraphics[angle=-90]{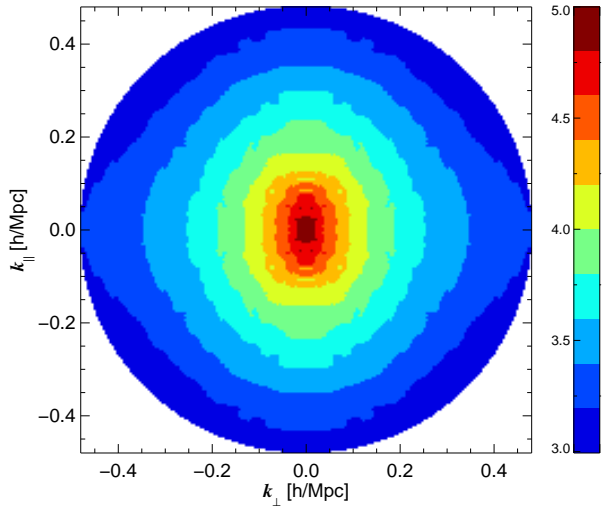}}
\caption{Illustration of the 2D power spectrum of galaxies in BOSS-DR11 
CMASS sample. We present a color map of the $\log_{10}P(\kp,\klos)$ as a 
function of $\kp$ and $\klos$. 
In the absence of RSD effect, we would obtain perfect circles with this 
discrete color scale. In the $\klos$ direction, we clearly see in the 
center the elongation due to the Kaiser effect, and in outskirts the 
squashing due to the FOG effect. }
\label{fig02}
\end{figure}

\begin{figure}
\resizebox{\hsize}{!}{\includegraphics{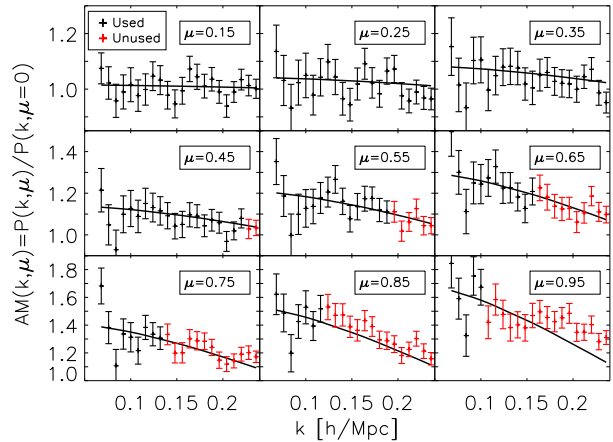}}
\caption{Power spectrum anisotropies of galaxies in BOSS-DR11 
CMASS sample. Each panel shows the anisotropic measure $\am(k,\mu)$ 
for one $\mu$ bin. Our measurements are indicated with the plus 
symbols and our best fit model (see \eqn\ref{EQ:Method:AMmodel}) with 
the solid lines. The errors are estimated using 1024 MD-Patchy mocks. 
Data points, with $k\mu<0.1\,h\mpc^{-1}$, used in the fitting are 
indicated in black, while unused data points are indicated in red.}
\label{fig03}
\end{figure}

\begin{figure*}
\centering
{\includegraphics[width=4.5in,angle=-90]{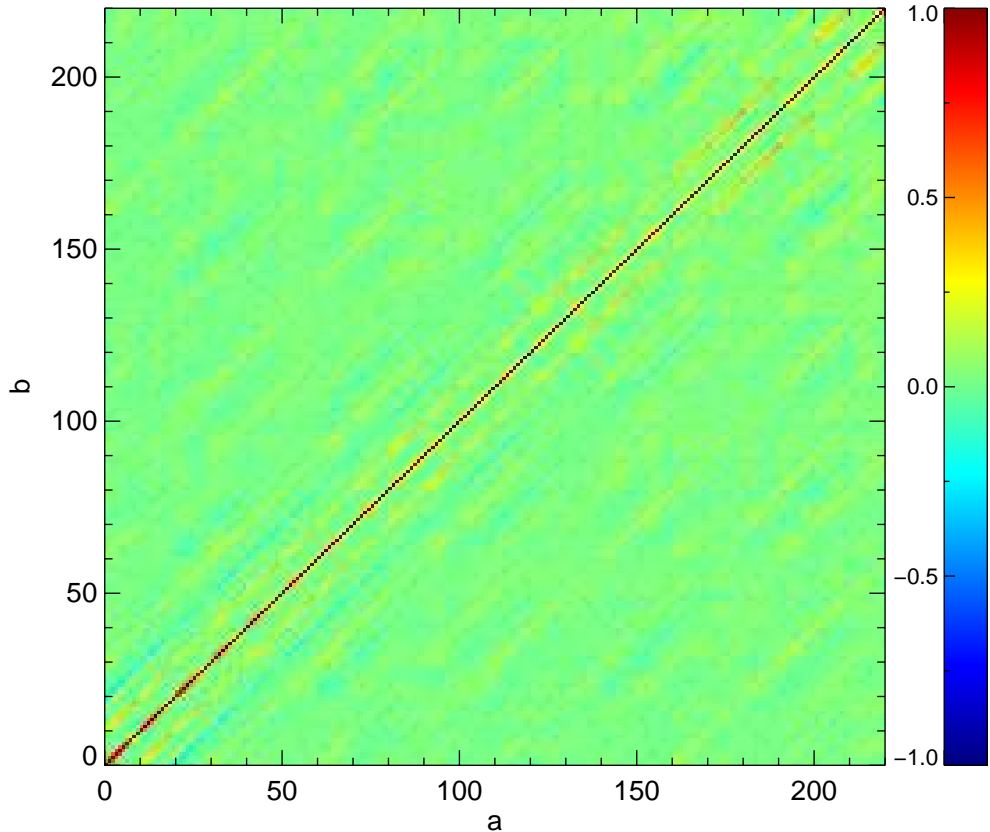}}
\caption{Correlation matrix of the 2D power spectrum for BOSS11-CMASS 
NGC galaxy sample, defined in \eqn\ref{EQ:2DPower:cov}. The colors
indicate correlation level. Red color stands for high correlation, 
blue for high anti-correlation and green for no-correlation. In the 
figure shows 220 modes in total ($10\,\mu\times22\,k$), 
that is $a,b=0,\dots,219$. }
\label{fig04}
\end{figure*}

\section{Results of Power Spectrum Measurement}
\label{Sec:res}
In \fig\ref{fig02}, we show the 2D power spectrum 
$P^s_g(\kp,\klos)$ for BOSS-DR11 CMASS galaxies, measured using our 
method. In order to highlight the features, we collect the value of 
the 2D power spectrum into 10 equal bins. The `elongated' feature in 
the central region is due to the large-scale (small $k$-value) Kaiser 
effect \citep{1987MNRAS.227....1K}. The `squashing' feature can also 
be seen in the outskirts,  where the FOG effect defeats the Kaiser 
effect. 

Although our method has been verified against mock galaxies in the 
$N$-body simulation (\S 2.2), we further perform two consistency 
tests against published BOSS measurements, to make sure that we 
have understood the BOSS data correctly and have applied our method 
correctly. These consistency tests are presented in the appendix.  
There we have shown that our measured 2D power spectrum can reproduce 
very well the correlation functions of  \cite{2014MNRAS.441...24A} 
and \cite{2014MNRAS.440.2692S}. This is not surprising by the
construction, but it does indicate that our measurement for the 2D
power spectrum is unbiased for the systematics such as the window 
function and shot noises. We have also shown that our 2D power 
spectrum, once convolved with the window function provided  by
\cite{2014MNRAS.443.1065B}, can very accurately reproduce their power
spectrum monopoles and quadrupoles as well, which further lends the 
support to our method.

One might worry about the possible coupling between different 
modes on the 2D power spectrum measurement since we rely on Fourier 
transforming the correlation function where the mode-coupling effect 
might be strong. In practice, we find that the resulting correlations 
between different modes using our method are weak. This can be 
understood since the window function has been decoupled in the 
measurement of 2D correlation function which reduces the mode-coupling 
effect. We show the correlation coefficients measured from 1024 
MD-Patchy mocks in \fig\ref{fig04}, which is defined as, 
\begin{eqnarray}
r^\mathrm{C}_{ab}=\frac{C_{ij,kl}}{\langle P^s_g(k_i,\mu_j)\rangle 
\langle P^s_g(k_k,\mu_l)\rangle}
\label{EQ:2DPower:cov}
\end{eqnarray}
The covariance matrix of the 2D power spectrum 
$C_{ij,kl}=\langle (P^s_g(k_i,\mu_j)-\bar{P}^s_g(k_i,\mu_j))
(P^s_g(k_k,\mu_l)-\bar{P}^s_g(k_k,\mu_l)) \rangle$ and 
$\bar{P}^s_g(k,\mu)=\langle P^s_g(k,\mu) \rangle$ is the mean power 
spectrum.  The bracket `$\langle \rangle$' represents the ensemble 
average, which is estimated using 1024 MD-Patchy mocks in this work.
The index $a,b$ is related to $k$-bin index and $\mu$-bin index as, 
$a=i\times N_\mu+j$ and $b=k\times N_\mu+l$, where 
$i,k=0,\dots,N_k-1$ and $j,l=0,\dots,N_\mu-1$. We have 10 $\mu$ bins 
($N_\mu=10$) and 22 $k$ bins ($N_k=22$) shown in \fig\ref{fig04}.
The size of $k$ bin is $\Delta k=0.008\mpch$.

The goal of 2D power spectrum measurement is to infer the information
of peculiar velocities. However, not all $P^s_g(\kp,\klos)$ measurement 
contain such useful information. Next we introduce a simple statistics, 
called the anisotropic measure (AM for short), to isolate the part 
useful for peculiar velocity and cosmology inference.  This is 
motivated by a neat feature of $P^s_g(\kp,\klos)$,  that 
$P^s_g(\kp,\klos=0)=P^s_g(k,\mu=0)$ is unaffected by RSD and equals the 
real-space power spectrum $P_g(k=\kp)$. Therefore all information of 
peculiar velocity is encoded in the ratio
\begin{equation}
\am(k,\mu)\equiv
\frac{P_g^s(k,\mu)}{P_g^s(k,\mu=0)}\ .
\label{EQ:Method:AM}
\end{equation}
AM describes the anisotropies of galaxy clustering induced by RSD. 
In contrast, the denominator $P_g^s(k,\mu=0)$ only contains the 
information of real-space clustering. Furthermore,  its modeling
is complicated by not only the nonlinear density evolution, but also
the scale dependence and nonlinearities in  galaxy bias. Therefore for
cosmological constraints from RSD, it is better to work on AM$(k,\mu)$ 
than on the full $P_g^s(k,\mu)$. In practice, we approximate the 
averaged $P^s_g(k,\mu)$ over $0.0<\mu<0.1$ as $P^s_g(k,\mu=0)=P_g(k)$, 
since for large $k$ of our interest, the RSD effect is negligible at 
$0.0<\mu<0.1$.

Since we Fourier transform correlation function to obtain 
$P^s_g(k,\mu)$ (and AM$(k,\mu)$), numerically we can get a continuous 
series of data points in $(k,\mu)$ space. However, the number of 
independent modes is limited by the survey volume and the value of 
$s_\mathrm{max}$.  The smallest dimension of CMASS sample is about 
$600 \mathrm{Mpc}$ ($z$ from 0.43 to 0.70), corresponding to 
$k\sim0.01\,h\mathrm{Mpc}^{-1}$. Moreover, the integral upper limit 
is adopted as $s_\mathrm{max}=300\,h^{-1}\mathrm{Mpc}$, which  sets 
a minimum value of $k$ to $\sim0.02\,h\mathrm{Mpc}^{-1}$. The 
signal-to-noise ratio on large scales is determined by cosmic 
variance (CV), $\mathrm{CV} \propto \sqrt{1/N(k,\mu)}$. Here 
$N(k,\mu)$ is the number of modes in the $(k,\mu)$ bin. This means 
that for $k$ close to $0.02\,h^{-1}\mathrm{Mpc}$, we do not have
information to further split into $\mu$ bins to infer RSD. Therefore,
in parameter fitting of 2D power spectrum in next section, we limit 
our analysis on scales of $k\ge0.064\,h/\mathrm{Mpc}$. We set the 
$k$-bin size to be $\Delta k=0.008\,h/\mathrm{Mpc}$ and $\mu$-bin 
size to be $\Delta\mu=0.1$.

%%%%%%%%%%%%%%%%%%%%%%%%%%%%%%%%%%%%%%%%%%%%%%%%%%%
%%%%%   Subsubsection: AM
%%%%%%%%%%%%%%%%%%%%%%%%%%%%%%%%%%%%%%%%%%%%%%%%%%%

We show AM$(k,\mu)$ for BOSS-DR11 CMASS galaxies in \fig\ref{fig03}, 
but  only data points with $k<0.24\,h\mpc^{-1}$ are shown.  Each panel 
shows $\am(k,\mu)$ for each $\mu$-bin.  For $\mu$ bins of small value, 
the RSD effect (AM$\neq 1$) is insignificant. But for $\mu$ bins of 
large value, the RSD effect is significant.  In particular we find 
that AM$(k,\mu)>1$ for all modes shown in \fig\ref{fig03}, 
meaning the Kaiser effect dominates over the FOG effect. But decreasing 
AM$(k,\mu)$ with increasing $k$ does show the increasing impact of FOG.

%%%%%%%%%%%%%%%%%%%%%%%%%%%%%%%%%%%%%%%%%%%%%%%%%%%%%%%%%%%%%%%%%%%%%%%%%%%%%%%%
%%%%%%%%%%%%%%%%%%%%%%%%%%%%%%%%%%%%%%%%%%%%%%%%%%%%%%%%%%%%%%%%%%%%%%%%%%%%%%%%
%%%%%%%%%%%%%%%%%%%%%%%%%%%%%%%%%%%%%%%%%%%%%%%%%%%%%%%%%%%%%%%%%%%%%%%%%%%%%%%%
%%%%%%%%%%%%%%%%%%%%%%%%%%%%%%%%%%%%%%%%%%%%%%%%%%%%%%%%%%%%%%%%%%%%%%%%%%%%%%%%
%%%%%%%%   Section: Growth rate   %%%%%%%%%%%%%%%%%%%
%%%%%%%%%%%%%%%%%%%%%%%%%%%%%%%%%%%%%%%%%%%%%%%%%%%%%%%%%%%%%%%%%%%%%%%%%%%%%%%%
%%%%%%%%%%%%%%%%%%%%%%%%%%%%%%%%%%%%%%%%%%%%%%%%%%%%%%%%%%%%%%%%%%%%%%%%%%%%%%%%
%%%%%%%%%%%%%%%%%%%%%%%%%%%%%%%%%%%%%%%%%%%%%%%%%%%%%%%%%%%%%%%%%%%%%%%%%%%%%%%%
%%%%%%%%%%%%%%%%%%%%%%%%%%%%%%%%%%%%%%%%%%%%%%%%%%%%%%%%%%%%%%%%%%%%%%%%%%%%%%%%

\section{Estimation of the structure growth rate}
\label{sec:cosmology}
Since RSD is induced by peculiar velocity and peculiar velocity is
related to matter density by the continuity equation, it allows us to
measure a specific combination of the structure growth rate, $fD$ (or
$f(z)\sigma_8(z)$). Here, $D$ is the linear density growth factor and
$f\equiv d\ln D/d\ln a$.  However, in reality it is highly nontrivial
to constrain $f(z)\sigma_8(z)$ from RSD, due to various complexities in
the RSD modeling \citep{2013PhRvD..87f3526Z}.  We do not aim to take
into account of all these complexities in our cosmological parameter
fitting. Since the nonlinearities are only important at large $k$ and
large $\mu$, it is possible to suppress their effects by selecting
data in the $(k,\mu)$ space where the quasi-linear approximation is
understood and works well. A data modeling including full
nonlinearities will be investigated in a future work.

%%%%%%%%%%%%%%%%%%%%%%%%%%%%%%%%%%%%%%%%%%%%%%%%%%%%%%%%%%%%%%%%%%%%%%%%%%%%%%%%
%%%%%%%%%%%%%%%%%%%%%%%%%%%%%%%%%%%%%%%%%%%%%%%%%%%%%%%%%%%%%%%%%%%%%%%%%%%%%%%%
%%%%%%%%   Subsection: Modelling   %%%%%%%%%%%%%%%%%%%
%%%%%%%%%%%%%%%%%%%%%%%%%%%%%%%%%%%%%%%%%%%%%%%%%%%%%%%%%%%%%%%%%%%%%%%%%%%%%%%%
%%%%%%%%%%%%%%%%%%%%%%%%%%%%%%%%%%%%%%%%%%%%%%%%%%%%%%%%%%%%%%%%%%%%%%%%%%%%%%%%

\subsection{RSD modeling} 
\label{Sec:Method:mod}
We adopt the RSD model constructed by \cite{2013PhRvD..87f3526Z}. The
redshift-space galaxy power spectrum can be written as, 
\begin{eqnarray}
P_g^s(k,\mu)&=& \{ P_g(k)(1+\beta W(k)\mu^2)^2+ 
   \mathrm{h.o.} \} \nonumber \\
   & & \times \exp\{-(k\mu\tilde{\sigma}_v)^2\}
\label{EQ:Method:pkmu}
\end{eqnarray}
Here $P_g(k)$ is galaxy power spectrum in real space. $\beta=f/b_g$ 
is the RSD parameter with $f$ the growth rate and $b_g$ the 
deterministic galaxy bias.  The FOG effect in this formalism has been 
derived to have a Gaussian form, with $\tilde{\sigma}_v$ the velocity 
dispersion in unit of $H(z)$.  The Gaussian form has been further 
verified in $N$-body simulations \citep{2013PhRvD..88j3510Z}. 

This formula contains two kinds of corrections to the commonly adopted
 Kaiser plus FOG formula.  The leading order correction is captured by 
$W(k)=P_{g\theta}(k)/\beta P_g(k)$. It  takes the nonlinear evolution 
of density-velocity relation into account and therefore the extends 
the Kaiser formula to nonlinear regime. $P_{g\theta}(k)$ is the 
density-velocity cross power spectrum and 
$\theta\equiv-\nabla\cdot\bfv/H(z)$. In the limit of large scales
density and velocity are perfectly correlated, so $W(k)=1$. 
Stochasticities develop in the density-velocity relation towards 
smaller scales and drives $W(k)<1$.  Eventually $W(k)\to0$ at deeply 
nonlinear region \citep{2013PhRvD..88j3510Z}. 

All high order corrections are collected into one term `$\mathrm{h.o.}$', 
with the exact expressions given in \cite{2013PhRvD..87f3526Z}. In 
principle, one would include all of the nonlinear terms to fit observed  
data. In this work, we take a different way to suppress the contribution 
of high-order terms by cutting data in $(k,\mu)$ space. This is motivated 
by the fact that we do not have a good understanding of nonlinearities 
and FOG effect which are very important to extracting the RSD parameter 
$\beta$ from galaxy clustering measurements. Although the FOG effect can 
be well described by a Gaussian damping function, the value of 
$\tilde{\sigma}_v$ is sensitive to how many nonlinear terms are included 
in the RSD models. Even though the full leading-order terms are included, 
the needed $\tilde{\sigma}_v$ for fitting the dark matter power spectrum 
still differs significantly from the velocity dispersion directly measured 
from the simulation \citep{2016arXiv160300101Z}. The authors of 
\cite{2016arXiv160300101Z} suggest that higher-order terms or including 
multi-streaming effect may explain the differences. This situation could 
be worse for galaxies since the galaxy bias will enhance the contribution 
of nonlinear terms. 

Good news is that the nonlinear corrections are important only at large 
$k$ and large $\mu$. If we restrict our analysis only at small $k$ and 
small $\mu$, the effect of high-order terms can be reduced greatly. 
Then the FOG effect and RSD parameter can be obtained faithfully. In 
this first analysis, we choose a strict limit on the available data, 
that is, we only include data points with $k\mu\le0.1\,h\mpc^{-1}$ 
and $k\le0.24\,h\mpc^{-1}$. This allows us to neglect the high-order
correction terms `h.o.' in \eqn\ref{EQ:Method:pkmu} 
\citep{2016arXiv160300101Z}.  We caution the readers that even with 
such strict cut, we still need to include $W(k)$, the effect of which is 
significant even at $k\sim 0.1h/$Mpc. We adopt a fitting formula for 
$W(k)$ given in \cite{2013PhRvD..88j3510Z}, 
\begin{eqnarray}
W(k)=1/[1+\Delta\alpha(z)\Delta_\mathrm{NL}^2(k,z)]
\label{EQ:Method:wk}
\end{eqnarray}
where $\Delta^2_\mathrm{NL}=k^3P_m(k)/2\pi^2$ is the nonlinear matter 
power spectrum. In practice, we use the mean value of $P^s_g(k,\mu=0)$ 
from 1024 mock samples divided by square of galaxy bias to estimate 
$P_m(k)$. The parameter $\Delta\alpha$ is also taken from 
\cite{2013PhRvD..88j3510Z} which is $0.376$ at redshift $z=0.5$.

Furthermore, instead of fitting against the measured $P_g^s(k,\mu)$, 
we fit against  the anisotropic measure defined in 
\eqn\ref{EQ:Method:AM} using the following model, 
\begin{equation}
\label{EQ:Method:AMmodel}
\amm(k,\mu)=(1+\beta W(k)\mu^2)^2\exp\{-(k\mu\tilde{\sigma}_v)^2\}\ .
\end{equation}
We use two parameters to model the anisotropic measure, the RSD parameter 
$\beta$ and reduced velocity dispersion $\tilde{\sigma}_v$. Recall that 
we use $P^s_g(k,\mu=0)$ to estimate the real-space galaxy power spectrum 
and to measure the galaxy bias $b_g\sigma_8(z)$. This frees us from modeling 
nonlinearities in the galaxy power spectrum. In the appendix \S 
\ref{Sec:Method:sys} , we show with  $N$-body simulation that neglecting 
high order corrections do not bias  the $\beta$ constraint, for the 
adopted cut of $k<0.24h/$Mpc and $k\mu\leq 0.1h/$Mpc.

%%%%%%%%%%%%%%%%%%%%%%%%%%%%%%%%%%%%%%%%%%%%%%%%%%%%%%%%%%%%%%%%%%%%%%%%%%%%%%%%
%%%%%%%%%%%%%%%%%%%%%%%%%%%%%%%%%%%%%%%%%%%%%%%%%%%%%%%%%%%%%%%%%%%%%%%%%%%%%%%%
%%%%%%%%   Subsection: Cov and like   %%%%%%%%%%%%%%%%%%%
%%%%%%%%%%%%%%%%%%%%%%%%%%%%%%%%%%%%%%%%%%%%%%%%%%%%%%%%%%%%%%%%%%%%%%%%%%%%%%%%
%%%%%%%%%%%%%%%%%%%%%%%%%%%%%%%%%%%%%%%%%%%%%%%%%%%%%%%%%%%%%%%%%%%%%%%%%%%%%%%%
\subsection{Covariance Matrix and Likelihood Analysis}
\label{Sec:Method:like}

The covariance matrix is estimated using 1024 MD-Patchy mocks, 
\begin{eqnarray}
C_{ij,kl}&=&\frac{1}{N_m-1}\sum_{n=1}^{N_m}(\am^{(n)}(k_i,\mu_j)-
            \widehat{\am}(k_i,\mu_j)) \nonumber \\
  & &\times(\am^{(n)}(k_k,\mu_l)-\widehat{\am}(k_k,\mu_l))
\label{Eq:Method:cov}
\end{eqnarray}
where $N_m=1024$ is the number of mock samples, $\am^{(n)}(k_i,\mu_j)$ 
is the anisotropic measure for the $n$-th mock and 
$\widehat{\am}(k_i,\mu_j)=\frac{1}{N_m}\sum_{n=1}^{N_m} \am^{(n)}(k_i,\mu_j)$. 

As shown by \cite{2007A&A...464..399H}, the inverse of the covariance 
matrix obtained above is a biased estimator of the true inverse 
covariance matrix. If the errors are Gaussian and data are 
statistically independent, we can rescale the obtained inverse 
covariance matrix by a constant factor to get unbiased inverse 
covariance matrix and log-likelihood function. The rescaling factor 
is $R_1=(N_m-N_d-2)/(N_m-1)$, where $N_d$ is the number of data points 
used in fitting procedure. 

The error of the obtained covariance matrix also should be taken into 
account to estimate parameters from it. We rescale the variance of 
parameters by another constant factor as proposed by 
\cite{2014MNRAS.439.2531P},  $R_2=\sqrt{(1+B(N_d-N_p)/(1+A+B(N_p+1))}$, 
where $A\approx2(N_m-N_d-2.5)^{-2}$ and $B\approx(N_m-N_d-3)^{-1}$ if 
$N_m-N_d \gg 1$.

\begin{figure}
\resizebox{\hsize}{!}
{\includegraphics{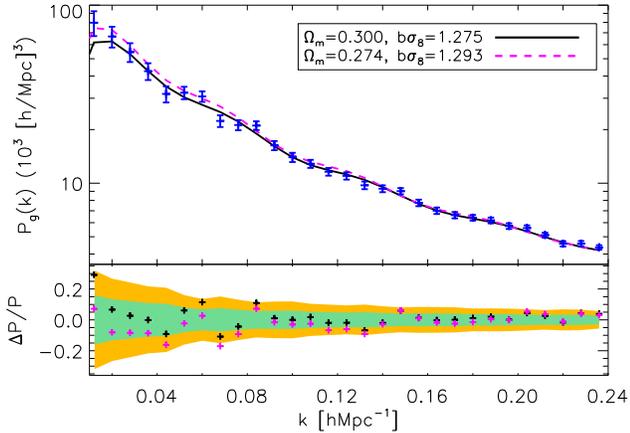}}
\caption{Real-space power spectrum of BOSS-DR11 CMASS galaxies. Upper 
panel: galaxy power spectrum in real space (black pluses with error bars 
estimated using 1024 MD-Patchy mocks, model power spectrum with bestfit 
$b\sigma_8(\zeff)$ for $\Omega_m=0.3$ (black solid line) and $\Omega_m=0.274$ 
(magenta dashed line) . Lower panel: the fractional difference between 
model and observation for $\Omega_m=0.3$ (black pluses) and 
$\Omega_m=0.274$ (magenta pluses). The Celadon green (Chrome yellow) 
band show the $1\sigma$ ($2\sigma$) level of the fractional difference.}
\label{fig05}
\end{figure}

We assume the noise of the anisotropic measure is Gaussian distributed 
and we estimate the covariance matrix using mock samples. Then we 
construct the likelihood function as, 
\begin{equation}
\mathscr{L} \propto \exp\{ - \chi^2(\mathbf{p},\mathbf{d})/2 \}
\end{equation}
where $\mathbf{p}$ is the parameter vector and $\mathbf{d}$ is data 
vector. The chi-square is defined as, 
\begin{eqnarray}
\chi^2(\mathbf{p},\mathbf{d})&=&\sum_{ij,kl}(\am(k_i,\mu_j)-\amm(k_i,\mu_j)) 
C^{-1}_{ij,kl} \nonumber \\
& & \times(\am(k_k,\mu_l)-\amm(k_k,\mu_l))
\end{eqnarray}
The model anisotropic measure $\amm(k_i,\mu_j)$ is given in
\eqn\ref{EQ:Method:AMmodel} with two parameters $\beta$ and 
$\tilde{\sigma}_v$. $C^{-1}_{ij,kl}$ is the inverse of 
covariance-matrix $C_{ij,kl}$ which is defined in 
\eqn\ref{Eq:Method:cov}.

As discussed in \seco\ref{Sec:Res:bias}, we can measure the galaxy 
bias factor $b_g\sigma_8(z)$ from the real-space power spectrum. 
Combining the measurement of RSD parameter $\beta$ and galaxy bias 
factor, we can calculate the posterior distribution function of the 
structure growth rate by 
\begin{eqnarray}
\mathcal{P}(f\sigma_8)=\int \mathcal{F}(b_g\sigma_8)\mathcal{G}
(\beta=\frac{f\sigma_8}{b\sigma_8}) \frac{d(b_g\sigma_8)}{b_g\sigma_8}\ .
\label{EQ:PDF:f}
\end{eqnarray}
Here $\mathcal{G}(\beta)$ is the posterior distribution function of 
RSD parameter $\beta$ and $\mathcal{F}(b_g\sigma_8)$ is the posterior 
distribution function of galaxy bias factor $b_g\sigma_8$. Notice 
that $f=f(z)$ and $\sigma_8=\sigma_8(z)$ in above equation.

The fiducial cosmology used to translate redshift to distance may 
differ from the true cosmology. This will introduce another distortion 
along the LOS direction and is called Alcock-Paczynski (AP) effect 
\citep{1979Natur.281..358A,2014ApJ...796..137L}. 
As an illustration of the methodology, we neglect the AP effect in 
this work. We leave the detailed modeling of the RSD effect and 
the joint-analysis of anisotropic measure and real-space galaxy power 
spectrum to a future work.

%%%%%%%%%%%%%%%%%%%%%%%%%%%%%%%%%%%%%%%%%%%%%%%%%%%%%%%%%%%%%%%%%%%%%%%%%%%%%%%%
%%%%%%%%%%%%%%%%%%%%%%%%%%%%%%%%%%%%%%%%%%%%%%%%%%%%%%%%%%%%%%%%%%%%%%%%%%%%%%%%
%%%%%%%%   Subsection: Growth rate   %%%%%%%%%%%%%%%%%%%
%%%%%%%%%%%%%%%%%%%%%%%%%%%%%%%%%%%%%%%%%%%%%%%%%%%%%%%%%%%%%%%%%%%%%%%%%%%%%%%%
%%%%%%%%%%%%%%%%%%%%%%%%%%%%%%%%%%%%%%%%%%%%%%%%%%%%%%%%%%%%%%%%%%%%%%%%%%%%%%%%
\begin{figure}
\resizebox{\hsize}{!}
{\includegraphics[angle=-90]{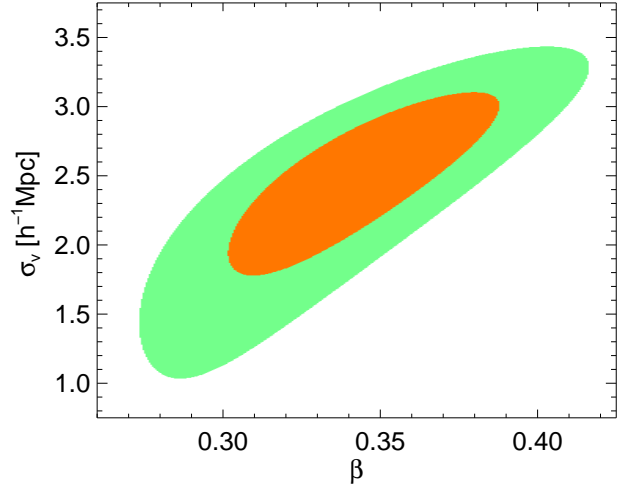}}
\caption{Two dimensional likelihood function of $\beta$ and $\tilde{\sigma}_v$ 
for BOSS-DR11 CMASS galaxies. The orange contours show 68\% confidence level, 
while the green contours show 95\% confidence level.}
\label{fig06}
\end{figure}

\subsection{Measuring $f\sigma_8(z)$}
\label{Sec:Method:growthrate}
We obtain the constraints on the structure growth rate in three steps, 
namely the measurement of galaxy bias $b_g\sigma_8(z)$, RSD parameter 
$\beta$ and growth rate $f(z)\sigma_8(z)$. 

%%%%%%%%%%%%%%%%%%%%%%%%%%%%%%%%%%%%%%%%%%%%%%%%%%%
%%%%%   Subsubsection: galaxy bias
%%%%%%%%%%%%%%%%%%%%%%%%%%%%%%%%%%%%%%%%%%%%%%%%%%%
\subsubsection{Galaxy Bias Factor}
\label{Sec:Res:bias}

We measure the combination of galaxy bias factor $b_g$ and 
$\sigma_8(z)$, which determines the amplitude of matter power spectrum, 
using the real-space galaxy power spectrum measurement $P_g(k)$. The 
real-space galaxy power spectrum can be obtained from the galaxy 2D 
power spectrum in the first $\mu$-bin $P_g(k)=P^s_g(k,\mu=0)$. 

In this work, we fix the shape of matter power spectrum and adopt a 
single-parameter model for the galaxy power spectrum in real space, 
\begin{eqnarray}
P_g(k)=\lambda^2\frac{P_{m,\mathrm{fid}}(k)}{\sigma_{8,\mathrm{fid}}^2(\zeff)}
\end{eqnarray}
where $\lambda=b_g\sigma_8(\zeff)$. We assume a $\lcdm$ 
cosmology for the fiducial matter power spectrum 
$P_{m,\mathrm{fid}}(k)$. Where the spectra index $n_s=0.96$ and 
Hubble parameter $h=0.7$. The baryon density parameter is given by 
$\Omega_b=\Omega_m/6$ and $\Omega_\Lambda=1-\Omega_m$. The matter 
power spectrum is calculated using `CAMB' \citep{2000ApJ...538..473L}. 
The nonlinear evolution is corrected using `halofit' model
\citep{2003MNRAS.341.1311S}. 
To study the systematics introduced by the choice of $\Omega_m$ in 
the measurement of $b_g\sigma_8(z)$, we construct two different fiducial 
matter power spectrum, one with $\Omega_m=0.274$ (close to the WMAP7 
bestfit value) and the other with $\Omega_m=0.3$ (close to the 
Planck2015 bestfit value). 

We obtain $b_g\sigma_8(\zeff)=1.293\pm0.007$ for $\Omega_m=0.274$ and 
$b_g\sigma_8(\zeff)=1.274\pm0.007$ for $\Omega_m=0.3$. This shows that 
the measurement of $b_g\sigma_8(\zeff)$ is insensitive to the value of 
$\Omega_m$ in this model. 
The bestfit model power spectrum and the measured real-space 
power spectrum are shown in \fig\ref{fig05}.

\begin{figure}
\resizebox{\hsize}{!}
{\includegraphics{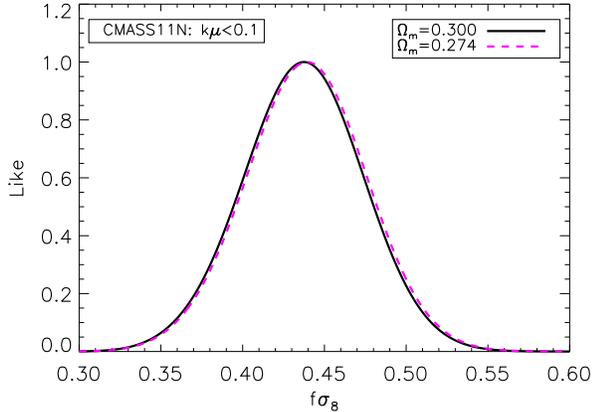}}
\caption{Normalized likelihood function of $f(\zeff)\sigma_8(\zeff)$ 
for BOSS-DR11 CMASS galaxies. The black solid curve shows results 
assuming $\Omega_m=0.3$ in measuring $b_g\sigma_8(\zeff)$ and the 
magenta dashed curve corresponds to $\Omega_m=0.274$. The figure 
shows negligible difference for these two cases. }
\label{fig07}
\end{figure}

%%%%%%%%%%%%%%%%%%%%%%%%%%%%%%%%%%%%%%%%%%%%%%%%%%%
%%%%%   Subsubsection: Beta
%%%%%%%%%%%%%%%%%%%%%%%%%%%%%%%%%%%%%%%%%%%%%%%%%%%
\subsubsection{Measurement of $\beta$}

Next, we do a maximum likelihood analysis to extract the RSD parameter 
$\beta$. The methods are discussed in \seco\ref{Sec:Method:mod}. We use 
two parameters - RSD parameter $\beta$ and reduced velocity dispersion 
$\tilde{\sigma}_v$, to model the theoretical anisotropic measure. In 
this paper, we fix the $W(k)$ function (\eqn\ref{EQ:Method:wk}) to 
reduce the number of parameters. The parameter $\Delta\alpha$ is 
calibrated in simulation \citep{2013PhRvD..88j3510Z}. We estimate 
the nonlinear matter power spectrum, which appears in $W(k)$, using 
the mean power spectrum of the 1024 MD-Patchy mocks. 

We calculate the likelihood function on grids of 2D parameter space 
and take 1000 points for each parameter placed equally over range of  
$0.1<\beta<0.6$ for RSD parameter and $0<\tilde{\sigma}_v<10$ for 
velocity dispersion. The normalized two-dimensional likelihood function 
of $\beta$ and $\tilde{\sigma}_v$ are shown in \fig\ref{fig06}.
The bestfit values are $\beta_\mathrm{best}=0.3448$ and 
$\tilde{\sigma}_{v,\mathrm{best}}=2.52\,h^{-1}\mpc$. We obtain the mean 
value and $1\,\sigma$ error of $\beta=0.3403\pm0.0285$ and 
$\tilde{\sigma}_v=2.40\pm0.44\,h^{-1}\mpc$, by marginalizing over 
$\tilde{\sigma}_v$ and $\beta$ respectively.

%%%%%%%%%%%%%%%%%%%%%%%%%%%%%%%%%%%%%%%%%%%%%%%%%%%
%%%%%   Subsubsection: Growth rate
%%%%%%%%%%%%%%%%%%%%%%%%%%%%%%%%%%%%%%%%%%%%%%%%%%%
\subsubsection{Measurement of $f(\zeff)\sigma_8(\zeff)$}
\label{Sec:Res:growth}

With the measurement of $\beta$ and $b_g\sigma_8(\zeff)$, we can 
obtain the growth rate of large scale structure $f(\zeff)\sigma_8(\zeff)$ 
through \eqn\ref{EQ:PDF:f}. We show the normalized likelihood function of
$f(\zeff)\sigma_8(\zeff)$ in \fig\ref{fig07}.  Same as $b\sigma_8(\zeff)$, 
the result $f(\zeff)\sigma_8(\zeff)$ is insensitive to the value of 
$\Omega_m$.  The bestfit value and $1\,\sigma$ error of the structure 
growth rate is $f(\zeff)\sigma_8(\zeff)=0.440\pm0.037$ for $\Omega_m=0.274$ 
and $f(\zeff)\sigma_8(\zeff)=0.438\pm0.037$ for $\Omega_m=0.3$.

%%%%%%%%%%%%%%%%%%%%%%%%%%%%%%%%%%%%%%%%%%%%%%%%%%%%%%%%%%%%%%%%%%%%%%%%%%%%%%%%
%%%%%%%%%%%%%%%%%%%%%%%%%%%%%%%%%%%%%%%%%%%%%%%%%%%%%%%%%%%%%%%%%%%%%%%%%%%%%%%%
%%%%%%%%%%%%%%%%%%%%%%%%%%%%%%%%%%%%%%%%%%%%%%%%%%%%%%%%%%%%%%%%%%%%%%%%%%%%%%%%
%%%%%%%%%%%%%%%%%%%%%%%%%%%%%%%%%%%%%%%%%%%%%%%%%%%%%%%%%%%%%%%%%%%%%%%%%%%%%%%%
%%%%%%%%   Section:  Discussion   %%%%%%%%%%%%%%%%%%%
%%%%%%%%%%%%%%%%%%%%%%%%%%%%%%%%%%%%%%%%%%%%%%%%%%%%%%%%%%%%%%%%%%%%%%%%%%%%%%%%
%%%%%%%%%%%%%%%%%%%%%%%%%%%%%%%%%%%%%%%%%%%%%%%%%%%%%%%%%%%%%%%%%%%%%%%%%%%%%%%%
%%%%%%%%%%%%%%%%%%%%%%%%%%%%%%%%%%%%%%%%%%%%%%%%%%%%%%%%%%%%%%%%%%%%%%%%%%%%%%%%
%%%%%%%%%%%%%%%%%%%%%%%%%%%%%%%%%%%%%%%%%%%%%%%%%%%%%%%%%%%%%%%%%%%%%%%%%%%%%%%%
\section{Discussions}
\label{Sec:Discuss}

\begin{figure}
\resizebox{\hsize}{!}
{\includegraphics{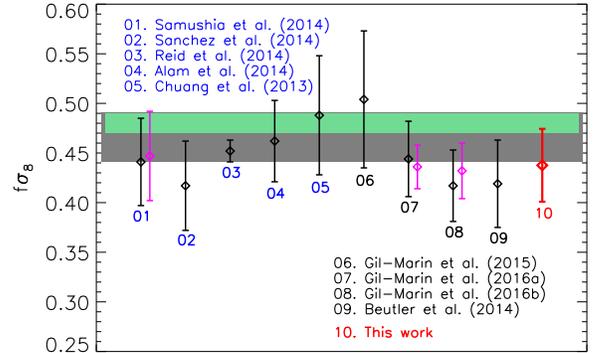}}
\caption{Constraints on $f(\zeff)\sigma_8(\zeff)$ from BOSS CMASS 
DR10, DR11 and DR12 release. Our result are shown in red diamond. 
Black diamonds show the results from various literatures. Magenta 
diamonds show those analysis that do not include the AP effect or 
use fiducial parameters for the AP effect. The green band show the 
$1\,\sigma$ confidence level allowed by Planck15 assuming $\lcdm$+GR 
model and grey band for WMAP9 assuming $\lcdm$+GR model.}
\label{fig08}
\end{figure}

In \fig\ref{fig08}, we present the reported measurements on 
$f(\zeff)\sigma_8(\zeff)$ at the same redshift of $\zeff=0.57$ 
\citep{2013arXiv1312.4889C,2014MNRAS.443.1065B,2014MNRAS.439.3504S,
2014MNRAS.440.2692S,2014MNRAS.444..476R,2015MNRAS.453.1754A,
2015MNRAS.451..539G,2016MNRAS.tmp..927G,2016arXiv160600439G}. 
Here we give a brief summary on the data, statistics, model and the 
measured $f(\zeff)\sigma_8(\zeff)$ of these analysis. Notice that 
(1)-(5) are using correlation function measurements and (6)-(9) are 
using power spectrum measurements.

\begin{enumerate}[(1)]

\item \cite{2014MNRAS.439.3504S} analyzed the monopole and quadrupole 
correlation function for BOSS-DR11 CMASS galaxies at scales of 
$24\,h^{-1}\mpc<s<152\,h^{-1}\mpc$. The model correlation function are 
calculated using the `streaming model'. They reported 
$f(\zeff)\sigma_8(\zeff)=0.441\pm0.044$ if the fitting method includes 
the AP effect and $f(\zeff)\sigma_8(\zeff)=0.447\pm0.028$ if the AP 
effect is fixed. 

\item \cite{2014MNRAS.440.2692S} analyzed the correlation function 
monopole and wedges of BOSS-DR11 CMASS galaxies for $s\le40\,h^{-1}\mpc$. 
Through their model based on the renormalized perturbation theory, they 
obtained $f(\zeff)\sigma_8(\zeff)=0.417\pm0.045$.

\item \cite{2014MNRAS.444..476R} explored the anisotropic clustering 
of BOSS-DR10 CMASS galaxies on small scales of $0.8\sim32\,h^{-1}\mpc$.
They reported a precise measurement of $f(\zeff)\sigma_8(\zeff)=0.450\pm0.011$.

\item \cite{2015MNRAS.453.1754A} analyzed the monopole and quadrupole 
correlation function for the BOSS-DR11 CMASS galaxies at scales of 
$30\sim126\,h^{-1}\mpc$. Based on the Convolution Lagrangian Perturbation 
Theory with Gaussian streaming model, they obtained 
$f(\zeff)\sigma_8(\zeff)=0.462\pm 0.041$.

\item \cite{2013arXiv1312.4889C} reported a detection of 
$f(\zeff)\sigma_8(\zeff)=0.488\pm0.060$ by analyzing the monopole and 
quadrupole correlation function of BOSS-DR12 CMASS galaxies at scales 
of $55<s<200\,h^{-1}\mpc$. 

\item \cite{2015MNRAS.451..539G} analyzed the monopole power spectrum 
and bispectrum of BOSS-DR11 CMASS galaxies at scales up to 
$k_\mathrm{max}=0.17\,h\mpc^{-1}$. They obtained 
$f(\zeff)^{0.43}\sigma_8(\zeff)=0.582\pm0.084$ without AP effect, 
which can be  transformed to $f(\zeff)\sigma_8(\zeff)=0.504\pm0.069$ 
by using the fiducial value of $f_\mathrm{fid}(\zeff)=0.777$.

\item \cite{2016MNRAS.tmp..927G} analyzed the monopole and quadrupole 
power spectrum of BOSS-DR12 CMASS galaxies at scales up to 
$k_\mathrm{max}=0.24\,h\mpc^{-1}$. The RSD effect are modeled based on 
\cite{2010PhRvD..82f3522T}. They obtained 
$f(\zeff)\sigma_8(\zeff)=0.444\pm 0.038$ if AP effect is included in 
the fitting method and $f(\zeff)\sigma_8(\zeff)=0.436\pm 0.022$ with 
fixed Hubble parameter and angular distance parameter. 

\item \cite{2016arXiv160600439G} improved the analysis of 
\cite{2015MNRAS.451..539G} by including more triangular shapes, using 
full covariance matrix, including quadrupole power spectrum and applying 
to BOSS-DR12 CMASS galaxies. Taking $k_\mathrm{max}=0.22\,h\mpc^{-1}$, 
they obtained $f(\zeff)\sigma_8(\zeff)=0.417\pm 0.027$ with AP effect 
included in the fitting method and $f(\zeff)\sigma_8(\zeff)=0.432\pm0.028$ 
with fixed AP effect.

\item \cite{2014MNRAS.443.1065B} measured the monopole and quadrupole 
power spectrum for BOSS-DR11 CMASS galaxies at scales up to
$k_\mathrm{max}=0.20\,h\mpc^{-1}$. Their power spectrum model was 
based on \cite{2010PhRvD..82f3522T} and the AP effect was included. 
They obtained $f(\zeff)\sigma_8(\zeff)=0.419\pm0.044$ and derived 
$\beta=0.342\pm0.037$.
\end{enumerate}

It's worth noting that our measurements of $f(\zeff)\sigma_8(\zeff)$ 
are consistent with \cite{2014MNRAS.439.3504S} at $1-2\%$ level. 
Although our measurements of $\beta$ are consistent with 
\cite{2014MNRAS.443.1065B} (within $1\%$), the values of 
$b\sigma_8(\zeff)$ are different by more than $4\%$ which results 
in a more than $4\%$ difference on $f(\zeff)\sigma_8(\zeff)$ 
measurement.

%%%%%%%%%%%%%%%%%%%%%%%%%%%%%%%%%%%%%%%%%%%%%%%%%%%%%%%%%%%%%%%%%%%%%%%%%%%%%%%%
%%%%%%%%%%%%%%%%%%%%%%%%%%%%%%%%%%%%%%%%%%%%%%%%%%%%%%%%%%%%%%%%%%%%%%%%%%%%%%%%
%%%%%%%%%%%%%%%%%%%%%%%%%%%%%%%%%%%%%%%%%%%%%%%%%%%%%%%%%%%%%%%%%%%%%%%%%%%%%%%%
%%%%%%%%%%%%%%%%%%%%%%%%%%%%%%%%%%%%%%%%%%%%%%%%%%%%%%%%%%%%%%%%%%%%%%%%%%%%%%%%
%%%%%%%%   Section:  Discussion   %%%%%%%%%%%%%%%%%%%
%%%%%%%%%%%%%%%%%%%%%%%%%%%%%%%%%%%%%%%%%%%%%%%%%%%%%%%%%%%%%%%%%%%%%%%%%%%%%%%%
%%%%%%%%%%%%%%%%%%%%%%%%%%%%%%%%%%%%%%%%%%%%%%%%%%%%%%%%%%%%%%%%%%%%%%%%%%%%%%%%
%%%%%%%%%%%%%%%%%%%%%%%%%%%%%%%%%%%%%%%%%%%%%%%%%%%%%%%%%%%%%%%%%%%%%%%%%%%%%%%%
%%%%%%%%%%%%%%%%%%%%%%%%%%%%%%%%%%%%%%%%%%%%%%%%%%%%%%%%%%%%%%%%%%%%%%%%%%%%%%%%
\section{Conclusion}
\label{Sec:Conclusion}

In this paper, we use the two-dimensional power spectrum 
in redshift space to measure the RSD effect. We revisit the method 
of measuring the galaxy 2D power spectrum by measuring and Fourier 
transforming the 2D correlation function. The 2D power spectrum 
measured in this way has several advantages:
\begin{inparaenum}[(A)]
\item they can improve the parallel-plane approximation and `moving-LOS' 
approximation and capture all RSD information under the assumption 
of distant observer and neglecting wide-angle effect; 
\item they are unbiased and free of normalization and shot-noise 
subtraction; 
\item the survey window function can be dealt with in configuration 
space; 
\item the nonuniform distribution of cosine angle $\mu$ can be solved.
\end{inparaenum}
Most importantly, working on 2D power spectrum opens the opportunity 
to separate the nonlinearities in the real-to-redshift space mapping 
at data level.

We have tested the 2D power spectrum measurements using mock galaxies 
constructed from high resolution CosmicGrowth $N$-body simulation and 
concluded that our method can give unbiased measurement of 2D power 
spectrum for large galaxy surveys. 
After applying the method on the BOSS-DR11 CMASS galaxy sample, we report 
for the first time the measurement of 2D power spectrum for this sample. 

We have introduced a new statistics, anisotropic measure, to extract
the structure growth rate from the 2D power spectrum measurements.  In
this paper we used a simple model with two parameters $\beta$ and
$\tilde{\sigma}_v$ to interpret this new measurement. We obtained
$\beta=0.3403\pm0.0285$ and $\tilde{\sigma}_v=2.40\pm0.44\,h^{-1}\mpc$
for BOSS-DR11 CMASS galaxies. We further measured the galaxy bias
factor from the real-space power spectrum, which is
$b_g\sigma_8(\zeff)=1.274\pm0.007$.  Combining the measurement of 
$\beta$ and $b_g\sigma_8(\zeff)$, we got the following measurement 
of the structure growth rate, 
$f(\zeff=0.57)\sigma_8(\zeff=0.57)=0.438\pm0.037$. This
measurement together with the 2D power spectrum can be used to put
interesting constraints on cosmological models. For this reason, we
will release our results of 2D power spectrum soon.

\section*{Acknowledgments}
This work was supported by the National Science Foundation of China
(Grants No. 11403071, 11320101002 \& 11533006) and National Basic 
Research Program of China (973 Programs No. 2015CB857001 \& 2015CB857003). 
ZL is supported by China Postdoctoral Science Foundation Funded Project 
(No. 2013M541511).

We acknowledge the BOSS collaboration to kindly release the BOSS DR11 
CMASS galaxy sample and the MD-Patchy mock sample plubicly.  Funding 
for SDSS-III has been provided by the Alfred P. Sloan Foundation, the 
Participating Institutions, the National Science Foundation, and the 
U.S. Department of Energy Office of Science. The SDSS-III web site is 
http://www.sdss3.org/.

%%%%%%%%%%%%%%%%%%%%%%%%%%%%%%%%%%%%%%%%%%%%%%%%%%%%%%%%%%%%%%%%%%%%%%%%%%%%%%%%
%%%%%%%%%%%%%%%%%%%%%%%%%%%%%%%%%%%%%%%%%%%%%%%%%%%%%%%%%%%%%%%%%%%%%%%%%%%%%%%%
%%%%%%%%%%%%%%%%%%%%%%%%%%%%%%%%%%%%%%%%%%%%%%%%%%%%%%%%%%%%%%%%%%%%%%%%%%%%%%%%
%%%%%%%%%%%%%%%%%%%%%%%%%%%%%%%%%%%%%%%%%%%%%%%%%%%%%%%%%%%%%%%%%%%%%%%%%%%%%%%%
%%%%%%%%   Appendix: consistency check   %%%%%%%%%%%%%%%%%%%
%%%%%%%%%%%%%%%%%%%%%%%%%%%%%%%%%%%%%%%%%%%%%%%%%%%%%%%%%%%%%%%%%%%%%%%%%%%%%%%%
%%%%%%%%%%%%%%%%%%%%%%%%%%%%%%%%%%%%%%%%%%%%%%%%%%%%%%%%%%%%%%%%%%%%%%%%%%%%%%%%
%%%%%%%%%%%%%%%%%%%%%%%%%%%%%%%%%%%%%%%%%%%%%%%%%%%%%%%%%%%%%%%%%%%%%%%%%%%%%%%%
%%%%%%%%%%%%%%%%%%%%%%%%%%%%%%%%%%%%%%%%%%%%%%%%%%%%%%%%%%%%%%%%%%%%%%%%%%%%%%%%

%\bibliographystyle{hapj}
%
%\bibliography{ms}

\appendix

%%%%%%%%%%%%%%%%%%%%%%%%%%%%%%%%%%%%%%%%%%%%%%%%%%%%%%%%%%%%%%%%%%%%%%
%%%%%% Section:  kmax for mock  %%%%%%%%%%%%%%%%%%%%%%%%%%%%%%%%%%%
%%%%%%%%%%%%%%%%%%%%%%%%%%%%%%%%%%%%%%%%%%%%%%%%%%%%%%%%%%%%%%%%%%%%%%
\section{Testing Systematics using N-body Simulation: value of $k_\mathrm{max}$}
\label{Sec:Method:sys}

The nonlinearities related to the real-to-redshift space mapping are 
important at large $\mu$ and large $k$ \citep{2016arXiv160300101Z}. 
In \seco\ref{Sec:Method:mod}, we mentioned that to suppress the effect 
of nonlinearities in the real-to-redshift space mapping, we would like 
to use data with the wavenumber $k$ and $k\mu$ smaller than some maximum 
value. Following the work of \cite{2016arXiv160300101Z}, we adopt 
$k\mu_\mathrm{max}=0.1\,h\mpc^{-1}$ for the BOSS-DR11 CMASS galaxies.
Now we determine the value of $k_\mathrm{max}$ using the mock galaxy 
catalog in the CosmicGrowth $N$-body simulation.

\begin{figure}
\centering
\includegraphics[width=4.5in]{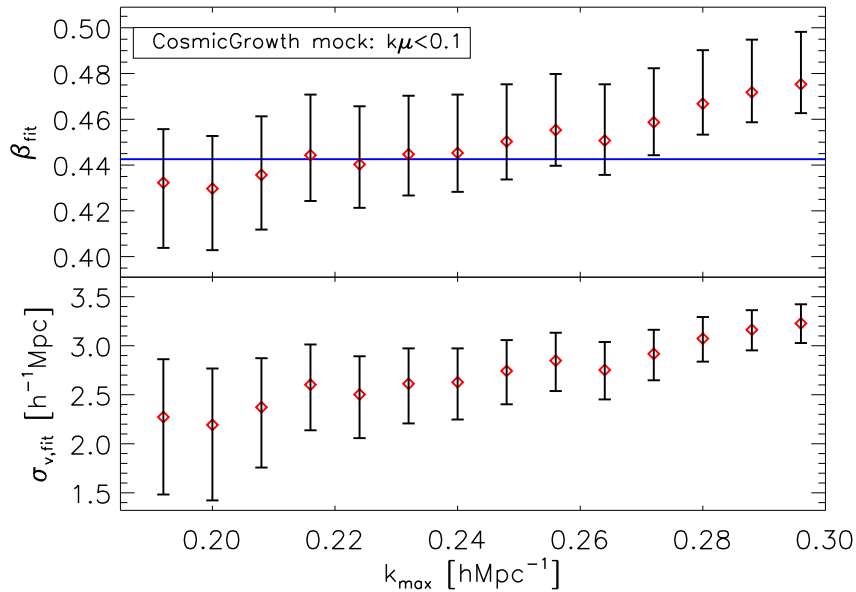}
\caption{Bestfit value of $\beta$ and $\tilde{\sigma}_v$ as a function 
of $k_\mathrm{max}$ for mock galaxies in CosmicGrowth simulation. 
Upper panel: red diamonds are bestfit values of $\beta$, black bars 
are the 68\%  confidence region of $\beta$ marginalizing over 
$\tilde{\sigma}_v$. Blue solid line shows the prediction of General 
Relativity with $\Omega_m=0.268$ and $b_\mathrm{gal}=1.68$.
Lower panel: same as upper panel, but for $\tilde{\sigma}_v$.}
\label{fig09}
\end{figure}

We obtained the anisotropic measure for the mock galaxy catalog 
following the procedure described in the main text. We assume a 
diagonal covariance matrix including the cosmic variance and shot 
noise. Next we do the likelihood analysis by using different values 
of $k_\mathrm{max}$. The bestfit value of $\beta$ and $\tilde{\sigma}_v$ 
as a function of $k_\mathrm{max}$ is shown in \fig\ref{fig09}. 
The $68\%$ confidence region of $\beta$ ($\tilde{\sigma}_v$) is
obtained by marginalizing over $\tilde{\sigma}_v$ ($\beta$). 
\fig\ref{fig09} suggests that $k_\mathrm{max}$ should be no 
more than $0.24\,h\mpc^{-1}$ to get unbiased measurement of $\beta$. 
Taking $k_\mathrm{max}$ greater than $0.24\,h\mpc^{-1}$ we will obtain 
positively biased estimate of $\beta$ which shows the nonlinear effect 
in real-to-redshift space mapping. Concerning the large statistical 
error, one would use $k_\mathrm{max}=0.27\,h\mpc^{-1}$ and the induced 
bias on $\beta$ is still within $1\sigma$ region of GR prediction. To 
be conservative,  we suggest to use $k_\mathrm{max}=0.24\,h\mpc^{-1}$. 
In the lower panel of \fig\ref{fig09}, we show the bestfit value 
of $\tilde{\sigma}_v$ as a function of $k_\mathrm{max}$, which is a
constant for our adopted $k_\mathrm{max}$

\begin{figure}
\centering
\includegraphics[width=5.0in]{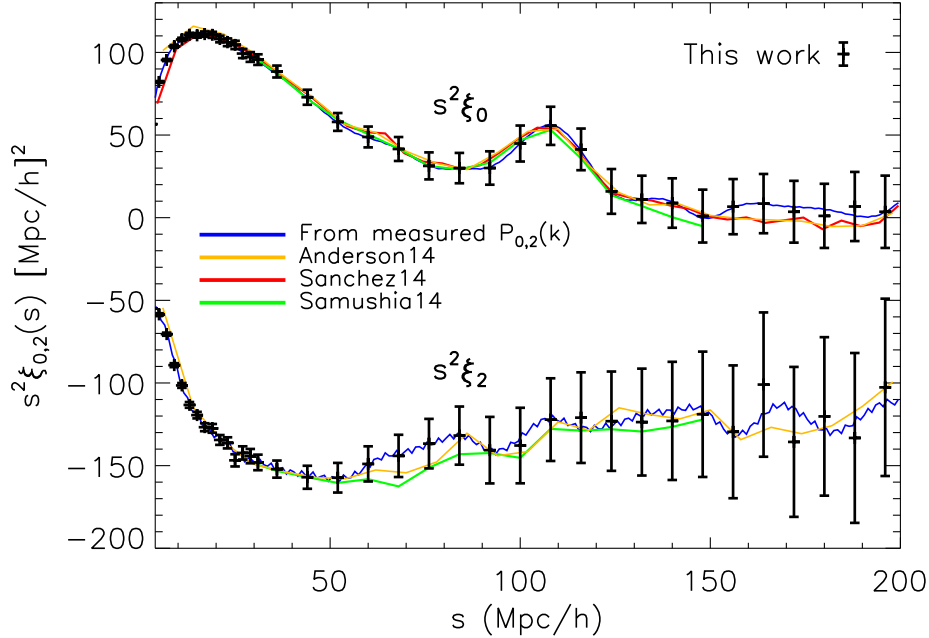}
\caption{Monopole and quadrupole correlation function of BOSS-DR11 
CMASS galaxies. Top of the figure show the monopole correlation 
function $s^2\xi_0(s)$, while bottom show quadrupole correlation 
function $s^s\xi_2(s)$ which have been shifted downward by a value 
of $80$. Black plus symbols represent our measurements 
using Landy-Szalay estimator. Errors are estimated using MD-Patchy 
mocks. Blue solid lines show the correlation function calculated 
from the measured power spectrum $P_{0,2}(k)$ using 
\eqn\ref{EQ:xifrompk}, respectively. We also show the measurements 
in literatures: brown solid lines for \cite{2014MNRAS.441...24A}, 
red solid lines for \cite{2014MNRAS.440.2692S} and blue solid lines 
for \cite{2014MNRAS.439.3504S}.}
\label{fig10}
\end{figure}

%%%%%%%%%%%%%%%%%%%%%%%%%%%%%%%%%%%%%%%%%%%%%%%%%%%%%%%%%%%%%%%%%%%%%%
%%%%%% Section: Comparison  %%%%%%%%%%%%%%%%%%%%%%%%%%%%%%%%%%%
%%%%%%%%%%%%%%%%%%%%%%%%%%%%%%%%%%%%%%%%%%%%%%%%%%%%%%%%%%%%%%%%%%%%%%
\section{Further consistent checks against existing BOSS measurements}
\label{Sec:appen:check}
Our correlation function based method of measuring the 2D power
spectrum has been verified against simulations. Here we show further
consistency tests. 

\subsection{Further consistent check: Recovered Correlation Function}
\label{Sec:appen:recf}
If our measured power spectrum is correct, with it one must be able to 
recover the measured correlation function by the BOSS collaborations. 
Therefore we calculate the monopole and quadrupole correlation from 
the power spectrum that we measured, by
\begin{eqnarray}
  \xi_l(s)= i^l \int P_l(k) j_l(ks) k^2 \frac{dk}{2\pi^2}\ .
\label{EQ:xifrompk}
\end{eqnarray}
Here $j_l(x)$ is the spherical Bessel function, $j_0(x)=\sin(x)/x$ 
and $j_2(x)=(3/x^2-1)\sin(x)/x-3\cos(x)/x^2$. We show them in 
\fig\ref{fig10}, against the measurements reported in 
literatures of \cite{2014MNRAS.441...24A}, 
\cite{2014MNRAS.439.3504S} and \cite{2014MNRAS.440.2692S}. 
The figure shows that we can successfully recover the monopole and 
quadrupole correlation function without any notable bias.

%%%%%%%%%%%%%%%%%%%%%%%%%%%%%%%%%%%%%%%%%%%%%%%%%%%
%%%%%   Subsubsection: multipole power spectrum
%%%%%%%%%%%%%%%%%%%%%%%%%%%%%%%%%%%%%%%%%%%%%%%%%%%
\subsection{Multipole Power Spectrum: {\rm CMASS} Galaxies}
\label{Sec:appen:pkl}

The BOSS collaboration has already measured the 2D power spectrum
monopole and quadrupole \citep{2014MNRAS.443.1065B,2014MNRAS.441...24A}. 
These measurements have not been corrected the window function effect. 
Therefore to compare with these results, we need to convolve the 
monopole and quadrupole that we measured, with the appropriate 
window function, 
\begin{equation}
  P^\mathrm{conv}_l(k)= 2\pi \sum_{l'} \int dk' k'^2 P_{l'}(k')
  |W(k,k')|_{ll'}^2\ .
\end{equation}
Here,  $|W(k,k')|^2_{ll'}$ is the BOSS-DR11 CMASS window function,
provided in \cite{2014MNRAS.443.1065B}. In light of the large 
effective volume of BOSS-DR11 CMASS sample, the window function only 
have small effect at very large scales and can be neglected at smaller 
scales. Here, we do not correct for the integral constraint effect, 
since both methods suffer from it.

\begin{figure}
\centering
\includegraphics[width=5.0in]{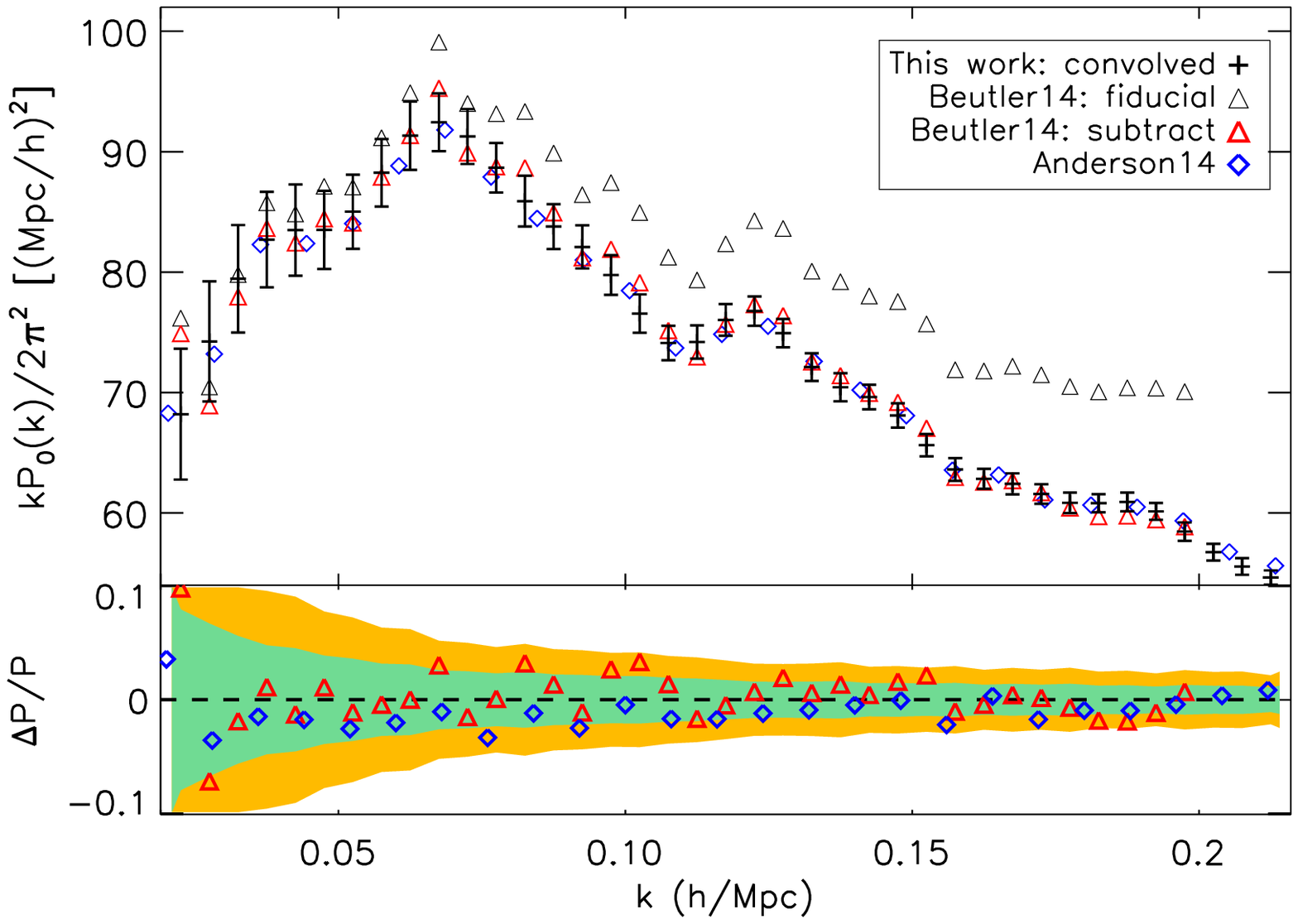}
\caption{Monopole power spectrum of BOSS-DR11 CMASS NGC galaxies. 
Top panel: monopole power spectrum, $kP_0(k)/2\pi^2$, in our analysis 
after being convolved with window function and rescaled by $\eta=0.90$ 
(black plus symbols), in \cite{2014MNRAS.443.1065B} (black triangles), 
in \cite{2014MNRAS.441...24A} (blue diamonds) and in 
\cite{2014MNRAS.443.1065B} after being subtracted a constant value of 
1120 on $P^\mathrm{B14}_0(k)$ (red triangles). Bottom panel: difference 
between our analysis and those in literatures, 
$\Delta P/P=(\tilde{P}_0(k)-P_0(k))/P_0(k)$ where 
$\tilde{P}_0(k)=P^\mathrm{B14}_0(k)-1120$ for 
\cite{2014MNRAS.443.1065B} and $\tilde{P}_0(k)=P^\mathrm{A14}_0(k)$ 
for \cite{2014MNRAS.441...24A}. 
Errors in our analysis are calculated using 1024 MD-Patchy mocks, 
also being convolved and rescaled in a way same as in observation.}
\label{fig11}
\end{figure}

\begin{figure}
\centering
\includegraphics[width=5.0in]{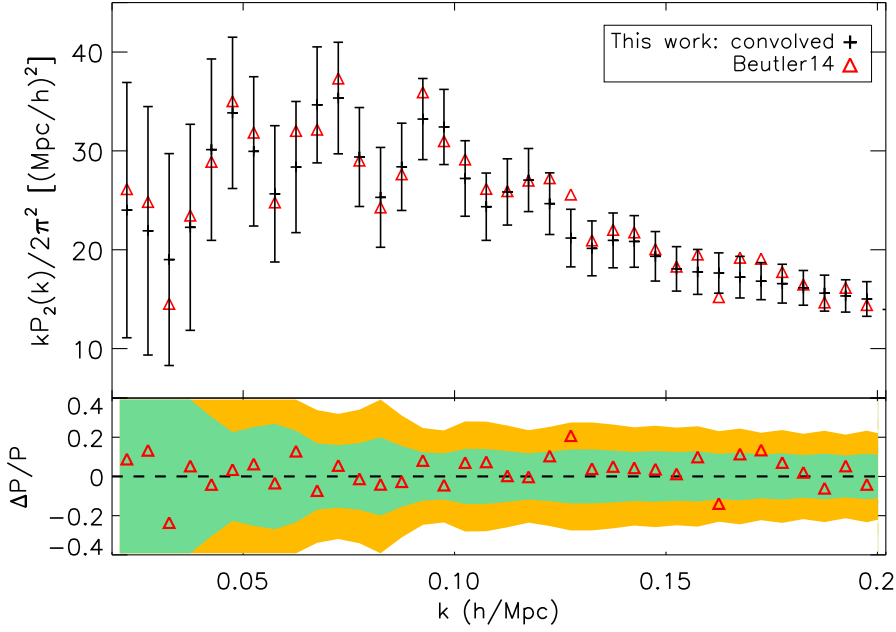}
\caption{Quadrupole power spectrum of BOSS-DR11 CMASS NGC galaxies. 
Top panel: quadrupole power spectrum, $kP_2(k)/2\pi^2$, in our 
analysis (black plus symbols) and in \cite{2014MNRAS.443.1065B} 
(red triangles). Bottom panel: difference between our analysis and 
those in \cite{2014MNRAS.443.1065B}, 
$\Delta P/P=(P^\mathrm{B14}_2(k)-P_2(k))/P_2(k)$. }
\label{fig12}
\end{figure}

The comparisons are shown in \fig\ref{fig11} for monopole 
power spectrum and in \fig\ref{fig12} for quadrupole power 
spectrum. We find that the convolved monopole power power spectrum 
match the measurements in \cite{2014MNRAS.443.1065B} and in 
\cite{2014MNRAS.441...24A} very well, up to a scaling factor $\eta=0.90$ 
($P^\mathrm{conv}_l(k)\rightarrow \eta P^\mathrm{conv}_l(k)$).  
Namely our power spectrum is about $10\%$ higher than theirs.  
Interestingly, by the same scaling, the convolved quadrupole power 
spectrum also match the measurements in \cite{2014MNRAS.443.1065B}.  
Such bias factor can result in about $4\%$ difference on the estimated 
$b_g\sigma_8$ and then on the growth rate $f\sigma_8$, although we 
obtained the same anisotropy measurement on the galaxy clustering as 
in \cite{2014MNRAS.443.1065B} or the same $\beta$ (\S \ref{sec:cosmology}).  
We do not know the origin of such bias at this
moment. We also notice that the monopole power spectra in
\cite{2014MNRAS.443.1065B} and \cite{2014MNRAS.441...24A} show
discrepancies too and differ by a shot-noise-like term of constant
amplitude. Nevertheless, our measurements of monopole/quadrupole power 
spectrum and monopole/quadrupole correlation function are 
self-consistent, as shown in previous subsection 
\S \ref{Sec:appen:recf}.

\end{document}